# Compressed Control of Complex Wireless Networks

Beatriz Lorenzo, *Member, IEEE* and Savo Glisic, *Senior Member, IEEE*

***Abstract*—Future wireless networks are envisioned to integrate multi-hop, multi-operator, multi-technology ($m^3$) components in order to meet the increasing traffic demand at an acceptable price for subscribers. The performance of such a network depends on the multitude of parameters defining traffic statistics, network topology/technology, channel characteristics and business models for multi-operator cooperation. So far, most of these aspects have been addressed separately in the literature. Since the above parameters are mutually dependent, and simultaneously present in a network, for a given channel and traffic statistics, a joint optimization of technology and business model parameters is required. In this paper, we present such joint models of complex wireless networks and introduce optimization with parameter clustering to solve the problem in a tractable way for large number of parameters. By parameter clustering we compress the optimization vector and significantly simplify system implementation, hence the algorithm is referred to as compressed control of wireless networks. Two distinct parameter compression techniques are introduced mainly parameter absorption and parameter aggregation. Numerical results obtained in this way demonstrate clear maximum in the network utility as a function of the network topology parameters. The results, for a specific network with traffic offloading, show that the cooperation decisions between the multiple operators will be significantly influenced by the traffic dynamics. For typical example scenarios, the optimum offloading price varies by factor 3 for different traffic patterns which justifies the use of dynamic strategies in the decision process. Besides, if user availability increases by multi-operator cooperation, network capacity can be increased up to 50% and network throughput up to 30-40%.**

## I. Introduction

THE vision of future wireless networks is evolving towards high density networks where multiple cellular network technologies such as 3G/4G/5G will coexist [1]. Recent studies predict that the explosive traffic growth will soon overload the cellular infrastructure resulting in poor performance or expensive service for subscribers [2]. To address this challenge, future wireless networks will integrate multi-hop, multi-operator, multi-technology ($m^3$) components in order to meet the increasing traffic demand at an acceptable price for subscribers. Regarding multi-hop architectures, there has been work on opportunistic networks [3], D2D [4], cognitive radio networks [5], more recently edge networks [4], etc. Multi-technology solutions for inter-working between cellular and WiFi [6] has attracted a lot of attention, and multiple operator schemes were addressed since the era of composite radio networks [7] until, more recently, on network sharing [8], [9]. Likewise there has been work on hierarchical cell structures. In the sequel, the most relevant solutions in each of these areas are reviewed.

Most of the works on multi-operator cooperation focus on spectrum sharing among different operators [8], [10]. Just since recently, multi-operator cooperation has been addressed in the context of cellular networks [9], [11]. In [9] an extensive business portfolio for wireless network operators is presented and authors discuss macro-economic aspects of multi-operator cooperative networks. In [11] a novel infrastructure sharing algorithm for multi-operator environments is presented which enables the deactivation of underutilized Base Stations (BSs) during low traffic periods. By using a game theoretic framework, mobile network operators individually estimate switching off probabilities to reduce their expected financial cost.

As cellular operators increase the coverage of their access networks it is more likely that there are overlaps which allows users to choose among multiple access opportunities. The issue of network selection or vertical handoff in a heterogeneous network has been extensively studied in the literature [12]. In [13], a market-based framework is developed in which a network selection mechanism is facilitated through first-price sealed-bid auction. Wireless network operators bid for the right to transport the subscriber's requested service over their infrastructure. The economic interaction between WiFi and WiMAX network providers is studied in [14]. A pricing model for bandwidth sharing in a WiMAX/WiFi network is presented and the optimal pricing solution is obtained by a Stackelberg leader-follower game. The economic incentive of the cellular operator to provide femtocell service is considered in [15]. They show that femtocell service can attract more users at a higher price and increase the operator´s profit.

However, end-users may not always be covered by any access point (AP) or may prefer shorter range transmissions. Thus, an efficient multi-hop routing protocol to identify the most appropriate AP and feasible relays in a multi-technology multi-operator network is needed. Multi-technology routing in heterogeneous networks is discussed in [16]-[18]. The authors in [16] addressed the importance of defining new metrics for routing decisions in heterogeneous networks. In [17] a hybrid proactive/reactive anycast routing protocol is proposed to discover the most suitable AP based on the path cost metrics, including hop count, energy cost, and traffic load. A WLAN-WiMAX routing protocol is developed in [18] where packet forwarding over the more stable WiMAX links is made in a topology-based manner, while position-based routing is exploited over WLAN links. The scheme also envisions the possibility of forwarding a packet over intermediate links of subscribers from other operators.

Most of the efforts in this area have addressed separately the issue of multi-operator cooperation [8], [10], multi-technology routing [16]-[18] and pricing models as incentives for cooperation among different networks [11], [13], [19]-[22]. In this paper, we present the way to optimize and control the network when all these components are present referred to as compressed optimization of complex networks as well as a unified model to analyze, in a tractable way, the impact of these solutions when used simultaneously in a complex network.

In our network model, the multi-hop concept is adopted in order to provide connectivity to the users that are not within direct coverage of any base station/access point. The reduced transmission range enabled by multi-hop makes the system transparent to mmWave technology which is anticipated to be used in the next generation of wireless networks. The transmission (relaying) range can be reduced to the point where the line of sight propagation dominates the received signal and the multipath components are completely eliminated. This range will be referred to as channel defading range.

The potential users acting as relays may belong to different operators and as such may or may not want to cooperate. In addition, multi-technologies are modeled by an assumption that some of the network subareas are also covered by femtocells or WLANs. Whenever available, these access points will offload the traffic from the cellular network and thus, contribute to the enhancement of the overall network capacity. For such network, new $m^3$ route discovery protocols are developed to find the most appropriate route towards the BS/AP and guarantee full connectivity within the network. After the most suitable routes are identified, the negotiation process between the multiple operators starts in order to reach a common access decision. A detailed analysis of cooperative multi-operator call/session access policies is presented. The policies are based on dynamic micro-economics of the multi-operators joint network access decisions. For a feasible implementation of the network optimization with such a multitude of parameters the model compression is used by introducing the network aggregation and network absorption functions. By using these techniques our network model can be optimized by acceptable complexity.

The contributions of this paper can be summarized as:

a) A new comprehensive model of multi-hop, multi-operator, multi-technology (m3) wireless networks, that enables a tractable analysis of the system, is presented.

b) New route discovery protocols for $m^3$ networks are proposed. These protocols are aware of users' availability to relay and mutual interference between all simultaneous routes in the network. An absorbing Markov chain is used for the analysis of the network where BS/APs are represented by absorbing states. The analysis also provides details on the complexity of finding the route towards the BS/AP and route delay as a function of relays' availability probability. This probability is obtained by parameter aggregation function that generates a new parameter representing a number of different phenomena in the network.

c) A new dynamic model of the joint decision process for traffic offloading between cellular and small service operators is proposed and analyzed. Our model quantifies the incentive for cooperation for each joint network/access decision based on dynamics of overall traffic in the network. As result, equilibrium price is obtained when offset of the utility (after and before offloading) for both operators is the same. This negotiation process leads to fair sharing of benefits in each joint access network decision.

d) The network optimization problem is defined to include a number of relevant parameters for $m^3$ networks such as: capacity, delay, power consumption on the route towards the BS/AP, users' availability and willingness to relay, multi-operator revenue, and offloading price. In the system optimization, model compression techniques are used based on network aggregation and absorption function.

A comprehensive set of numerical results is presented to show the impact of the offloading decision on the network performance. The performance of the $m^3$ routing protocols is shown in terms of the throughput, delay, power consumption and complexity where different sets of users are unavailable to relay.

The rest of the paper is organized as follows. In Section II we introduce the system model and notation. Network model compression is described in Section III. The $m^3$ route discovery protocols are presented in Section IV. Performance analysis is given in Section V whereas the traffic offloading incentives are presented in Section VI. Performance evaluation and implementation details are given in Section VII, and Section VIII concludes the paper.

## II. System Model and Notation

In this section, we describe the system model that integrates multi-hop, multi-operator and multi-technology components as shown in Fig. 1. Each of these components will be explained in the following subsections.

### A. *Multi-hop*

Multi-hop transmission is modeled by considering the cell tessellation scheme presented in [23], where the macrocell of radius $R$ is formally divided into inner hexagonal subcells[1] of radius $r < R$ as shown in Fig.1. This models the relative positions of the potential relays rather than the physical existence of the subcells. We consider uplink transmission and uniform distribution of the users across the cell. It is assumed that a potential, ready to cooperate, transmitter/receiver is on average situated in the center of each subcell. So, the users transmit uplink by relaying to adjacent users on the way to the BS. If a user is unavailable to relay, it may be because of lack of coverage, limited battery life, or belonging to a different operator with no mutual agreement for cooperation. The last case will be elaborated in detail in the next subsection.

[1] Suppose that all terminals transmit with the same power.

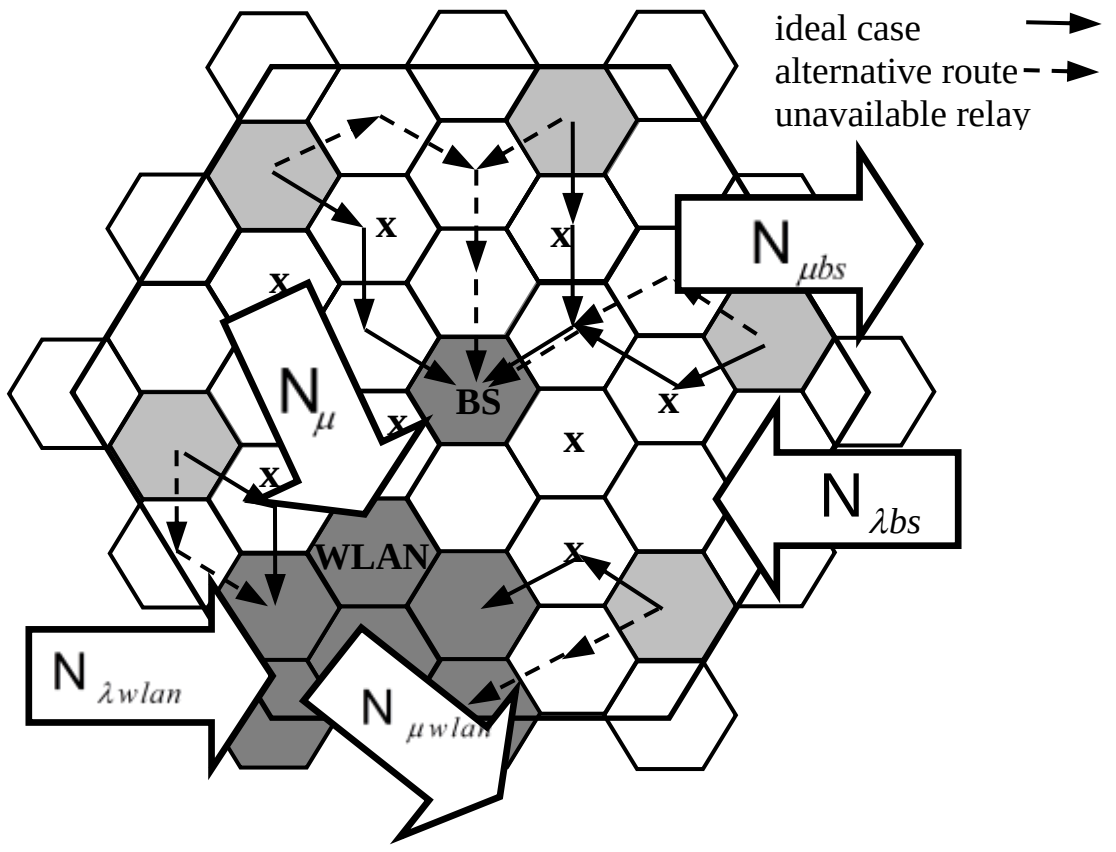

Fig 1. $m^3$ wireless network.

The BS is surrounded by $H$ concentric rings of subcells. For the example in Fig.1, $H$=4. Let us denote the location of the user by using polar coordinates as $u_{h,\theta}$ where $h$ is the ring index, $h$ = 1,…, $H$ and $\theta$ is the angle. We assume that user $u_{h,\theta}$ is transmitting to $u_{h',\theta'}$ and a cochannel interfering user $u_{\eta,\varphi}$ transmits simultaneously. Then, the interference distance which is the distance between the interfering user $u_{\eta,\varphi}$ and the referent receiver $u_{h',\theta'}$ denoted as $d_{\eta,\varphi;h',\theta'}$ can be obtained by the cosine theorem as $d_{\eta,\varphi;h',\theta'} = d_r\sqrt{(h')^2+(\eta)^2-2\cdot(h')\cdot\eta\cdot\cos(\theta',\varphi)} = d_r\cdot Z_{\eta,\varphi;h',\theta'}$ where $d_r$ is the relaying distance, $d_r = d_{h,\theta;h',\theta'}$. The signal to interference plus noise ratio SINR at $u_{h',\theta'}$ is defined as [23]

$$SINR_{h',\theta'} = G_{h,h'}P \Big/ \left(\sum_{\eta,\varphi} I_{\eta,\varphi;h',\theta'} + N_{h'}\right) = \frac{P}{\sum_{\eta,\varphi} P / Z^{\alpha}_{\eta,\varphi;h',\theta'} + \xi_{h'}\cdot(\sqrt{3}\cdot r)^{\alpha}} \tag{1}$$

where $P$ is the transmission power, $G_{h,h'}$ is the channel gain between $u_{h,\theta}$ and $u_{h',\theta'}$, $I_{\eta,\varphi;h',\theta'} = G_{\eta,\varphi;h',\theta'}P$ is the interference, $G_{\eta,\varphi;h',\theta'}$ is the channel gain between cochannel interfering user $u_{\eta,\varphi}$ and the referent receiver $u_{h',\theta'}$, $d_r = \sqrt{3}\cdot r$, $\alpha$ is the propagation constant and $\xi_{h'}$ is the background noise power. As we consider a dense network, the channel model considered includes the propagation losses, but not the effects of fading due to the proximity between the users as in [23]. The Shannon channel capacity is obtained as $c_l = \log(1+SINR_{h',\theta'})$. It is worth noticing that (1) can be written as a function of $H$ given that $H = \lfloor R/(2r) \rfloor$. Cell/subcell geometry relations will be used in the next section to define parameter absorption technique that enables model compression.

Moreover, the reduced transmission range in multi-hop concept makes the system transparent to the use of mmWave technology which is a promising technology for future cellular networks. For the application of mmWave, the physical layer model should be readjusted accordingly [24]. One of the parameters relevant for this analysis is the antenna beamwidth $\varphi$ by which a terminal will be visible by its neighbors with probability $p_\varphi = \varphi$ / 360.

### *B. Multiple Cellular Network Operators Cooperation*

We model the scenario where a number of operators coexist in the cellular network. It is assumed that a single operator $i$ has a terminal available in a given subcell with probability $p_{o_i}$. In a multi-operator cooperative network, there will be a terminal available for relaying in the same subcell if at least one out of $N_0$ operators has a terminal at that location. This will occur with probability

$$p = 1 - \prod_{i=1}^{N_0}(1-p_{o_i}) \tag{2}$$

This relation will be used in next section as a basis to introduce parameter aggregation technique for model compression. This probability is higher for higher number of operators willing to cooperate. In general, this will result into a reduction of the relaying route length which is illustrated in Fig. 1 where the ideal case refers to full cooperation ($p$ = 1). If operators cooperate and let their users to flexibly connect to the BS that is more convenient to them, the capacity of both operators will improve. Thus, a better performance of the network will be obtained in the multi-operator cooperative scenario as shown in Section VI. Otherwise, if there is no willingness to cooperate, alternative routes will be used as shown in Fig.1 in dashed lines.

### *C. Multiple Operators Cooperation with Multiple Technologies*

In general multiple technologies will be available in a heterogeneous network which enables more appropriate AP choice at a specific place and time based on users' requirements.

In this subsection, we model the scenario where the cellular network operator is interested in cooperating with a WLAN operator to offload some of its users through a WLAN. Similar relation can be established with a small cell owner. Fig.1 shows this scenario where the cellular network is overlapping in coverage with a WLAN, presented as a cluster of 6 subcells in the lower left corner of the cell. It is assumed that the WLAN uses different channels than the macrocell, so there is not interference among those links. As result, independent scheduling will be performed in both networks. The capacity of the WLAN´s links is obtained as in Section II.A by considering now that users $u_{h,\theta}$, $u_{h',\theta'}$, and $u_{\eta,\varphi}$ belong to the WLAN.

As we can see in Fig. 1, if cellular and WLAN operators cooperate, cellular users located close to the WLAN could be offloaded through that network. Consequently, the new routes will be shorter and in general it will result in shorter scheduling interval. If the number of users currently served by the BS is large and the WLAN is not overloaded, a reasonable price will be charged for offloading and thus, both networks will benefit.

### E. Notation

In order to model traffic dynamics and offloading process shown in Fig. 1, the following notation will be used through the paper.

We denote by $\mathsf{N}_{bs}$ and $\mathsf{N}_{wlan}$ the set of users transmitting to the BS and WLAN, respectively. The set of new users arriving at a given instant to the macrocell and WLAN are denoted as $\mathsf{N}_{\lambda bs}$ and $\mathsf{N}_{\lambda wlan}$, respectively. The set of users leaving each network (session terminated) at a given instant is denoted as $\mathsf{N}_{\mu bs}$ and $\mathsf{N}_{\mu wlan}$. The set of users handed off from the macrocell to the WLAN at the given instant is denoted by $\mathsf{N}_{\mu}$.

In the subsequent time instant ($t^+$) when the offloading decision has been made, the set of users connected to the BS, $\mathsf{N}_{bs}^{+}$, and WLAN, $\mathsf{N}_{wlan}^{+}$, can be represented as

$$\mathsf{N}_{bs}^{+} = \mathsf{N}_{bs} \cup \mathsf{N}_{\lambda bs} \setminus \mathsf{N}_{\mu bs} \setminus \mathsf{N}_{\mu} \tag{3a}$$

$$\mathsf{N}_{wlan}^{+} = \mathsf{N}_{wlan} \cup \mathsf{N}_{\lambda wlan} \setminus \mathsf{N}_{\mu wlan} \cup \mathsf{N}_{\mu} \tag{3b}$$

## III. NETWORK MODEL COMPRESSION

In this section we present the network model compression for feasible implementation and optimization of the network where all components described in Section II coexist. The *network aggregation* function $f_{ag}(\mathcal{A}/\mathcal{B})$ replaces the set of parameters $\mathcal{A}$ with a new, smaller set of parameters $\mathcal{B}$ that has an equivalent impact on the system utility function with $\mathcal{A} \supseteq \mathcal{B}$. The *network absorption* function $f_{ab}(\mathcal{A} \cup \mathcal{B} \to \mathcal{A})$ represents parameters from set $\mathcal{B}$ as a function of parameters from set $\mathcal{A}$ so that resulting set size of $\mathcal{A} \cup \mathcal{B}$ is the same as the size of set $\mathcal{A}$ (i.e. $\mathcal{B}$ is absorbed by $\mathcal{A}$).

On a higher level of abstraction the network from the previous section can be represented by its state vector where each component is defined as in Table I. If we control the network topology by changing parameter $H$, a number of network state vector components can be absorbed by this control parameter as shown in Table II. Given the geometry of the cell, parameter $N$ is absorbed by $H$ and $\mathbf{p}_o$ is absorbed by $\mathbf{n}_o$. Similarly, the terminal availability probability $p_a$ is absorbed by traffic parameters $\zeta = \lambda / \mu$, where $\lambda$ is the arrival rate and $\mu$ the service rate. The probability of terminal being seen $p_\varphi$ is absorbed by antenna beam width $\varphi$. Then, the aggregation function $p = 1 - \prod_{i=1}^{N_0} (1 - p_a p_\varphi p_{o_i})$ replaces probabilities $p_a$, $p_\varphi$ and $p_{oi}$ with the overall availability probability $p \rightarrow f_{ag}(\mathcal{A}/\mathcal{B}) \rightarrow f_{ag}(p_a, p_\varphi, p_{o_i} / p)$. Similarly, the channel gain and interference are absorbed by $H$ and, $H$ and $P$, respectively. Finally, parameter $w$ (terminal reward for relaying) is absorbed as well by $\zeta$. The price for relaying is higher if terminals' own traffic is higher (higher arrival rate and thus, higher $\zeta$) and $\gamma$ is a proportionality constant. The optimization problem will now have one optimization variable $H$ and only four system parameters $\mathbf{n}_o$, $p$, $\zeta$ and $\varphi$. By using this high level abstraction network model, elaborating any specific case is straightforward.

Table I. Network state vector $\boldsymbol{v}$ ($H, N, \mathbf{n}_o, \mathbf{p}_o, p_a, G, I, \zeta, \varphi, p_\varphi, w$) $\in \mathbb{V}$

| | |
|---|---|
| $H$<br>$N$ | Number of rings<br>Number of subcells |
| $n_{o_i}$, $\mathbf{n}_o = n_{o_i}$ | Number of terminals of operator $i$ in a cell |
| $p_{o_i}$, $\mathbf{p}_o = p_{o_i}$<br>$p_{o_i} = n_{o_i} / N$ | Probability that a single operator $i$ has a terminal in a given subcell |
| $p_a$ | Probability that the given terminal is available for relaying |
| $G$ | Channel gain |
| $I$ | Interference |
| $\zeta = \lambda / \mu$ | Call arrival/service ratio per terminal |
| $\varphi$<br>$p_\varphi$ | Antenna beam width<br>Probability of terminal being seen |
| $w$ | Award for a terminal to serve as a relay |

Table II. Compressed network state vector $\boldsymbol{v}^{(c)}$ ($H, \mathbf{n}_o, p, \zeta, \varphi$) $\in \mathbb{V}^{(c)}$

| | |
|---|---|
| $H$<br>$N = 3H(H+1)$ | Absorption ($N$ represented /absorbed by $H$) |
| $\mathbf{n}_o = n_{o_i}$<br>$\mathbf{p}_o = p_{o_i}$<br>$p_{o_i} = n_{o_i} / N = n_{o_i} / (3H(H+1))$ | Absorption<br>$p_{o_i}$ represented by $n_{o_i}$ and $N$ by $H$ |
| $\zeta = \lambda / \mu$<br>$p_a = f_{ab}(p_a, \zeta \longrightarrow \zeta) = 1 - \zeta$ | Absorption ( $p_a$ represented by $\zeta$) |
| $\varphi$<br>$p_\varphi = f_{ab}(p_\varphi, \varphi \to \varphi) = \varphi / 360$ | Absorption ($p_\varphi$ represented by $\varphi$) |
| $p = f_{ag}(\mathbf{p}_o, p_a, p_\varphi / p)$<br>$p = 1 - \prod_{i=1}^{N_0} (1 - p_a p_\varphi p_{o_i})$ | Probability that at least one terminal is available for relaying Aggregation ($p_a$, $p_\varphi$, $p_{oi}$ represented by $p$) |
| $G = f_{ab}(G, H \to H) \approx 1/(d_r)^\alpha$<br>$\approx \left(2H / (\sqrt{3}R)\right)^\alpha$ | Absorption ($G$ represented by $H$) |
| $I = f_{ab}(I, H, P \longrightarrow H, P)$<br>$I = \sum_{\eta,\varphi} I_{\eta,\varphi;h',\theta'}$<br>$= \sum_{\eta,\varphi} \left(2H / (\sqrt{3}R \cdot Z_{\eta,\varphi;h',\theta'})\right)^\alpha P$ | Absorption ($I$ represented by $H$ and $P$=*constant*) |
| $w = f_{ab}(w, \zeta \longrightarrow \zeta) = \gamma \cdot \zeta$ | Absorption ($w$ represented by $\zeta$, $\gamma$=*constant*) |

In the compression process (aggregation/absorption) a parameter can be excluded from the optimization vector if it can be represented equivalently by another parameter already existing in the optimization vector. The previous model compression enables simple network topology control as

$$H \Leftarrow \begin{cases} H+1, & \text{if } U(H+1) > U(H) \\ H-1, & \text{if } U(H-1) > U(H) \\ H, & \text{otherwise} \end{cases} \tag{4}$$

where $U$ is the system utility function. If the utility of the topology with $H$+1 hops is larger than for the case of $H$ hops, the network topology will be updated to $H$+1 and vice versa.

## IV. $m^3$ ROUTE DISCOVERY PROTOCOLS

In this section, we present two route discovery protocols for $m^3$ wireless networks. Then we model these protocols by using model compression techniques introduced in Section III. Multi-hop routing is used to establish a route for those users that are not directly covered by any AP. It can also optimize power consumption in the network. The protocols are intended for the situation where some of the users are not available to relay due to lack of coverage, interference, or noncooperation between different operators. Later on, the best route in terms of the given utility is chosen. Details on protocol implementation are provided in Section VII.D.

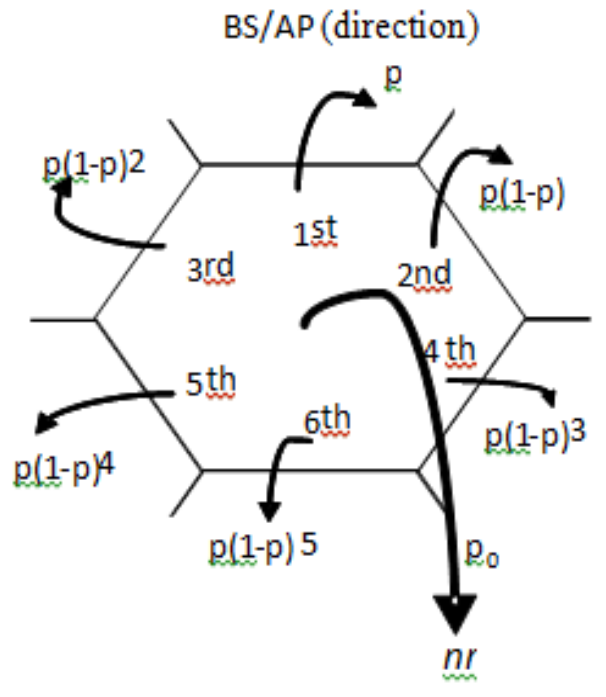

Fig. 2. Relaying alternatives for MDR.

### A. *Minimum Distance Routing (MDR)*

In general, we assume that the order in which this protocol tries the possible relaying alternatives is given in Fig.2. First, the protocol checks the adjacent user that is in the direction with *the shortest distance towards the BS/AP*. The user will be available with probability $p$ as shown in Fig.2 and if available, relaying will take place as indicated. Parameter $p$ is obtained by aggregation process discussed in Section III. If this user is not available, then the protocol checks the availability of the next user in the order indicated in Fig.2. First, it checks the right user, which will be available with probability $p$, so the probability that this transition will take place is $p(1-p)$. In the case of non availability the protocol will check the left user. The protocol continues in the same fashion until it gets to the last adjacent user, and relaying will take place with probability $p(1-p)^5$. If none of the above options is available, then the route will not be established with probability $p_0$ as indicated in Fig. 2, where *nr* refers to no route state.

In order to avoid excessive deviations in the length of the route, the number of possible relaying alternatives for a given node can be limited to $K'$. For the tessellation scheme used in Fig. 1, $K' = 6$. Once all the routes are found, the transmissions are scheduled in different time slots. One option is to let the users transmit in the same slot for as long as there is no collision in the transmission. Conventional or soft graph colouring techniques [25] can be used to optimize the subsets of users allowed to transmit simultaneously. As the search for the optimum scheduling in a multi-hop network is a NP-hard problem, we suggest the following alternative which is straightforward for practical implementation.

We apply a conventional resource reuse scheme used for cellular networks to our tessellation scheme, as shown in Fig. 3, for the resource reuse factor $K = 7$. The clustering factor $K$, equivalent to the frequency reuse factor in cellular networks, partitions the network into clusters of $K$ different types of users. The type of user $k$ is determined by its position within the cluster ($k = 1, 2, 3,\dots, K$).

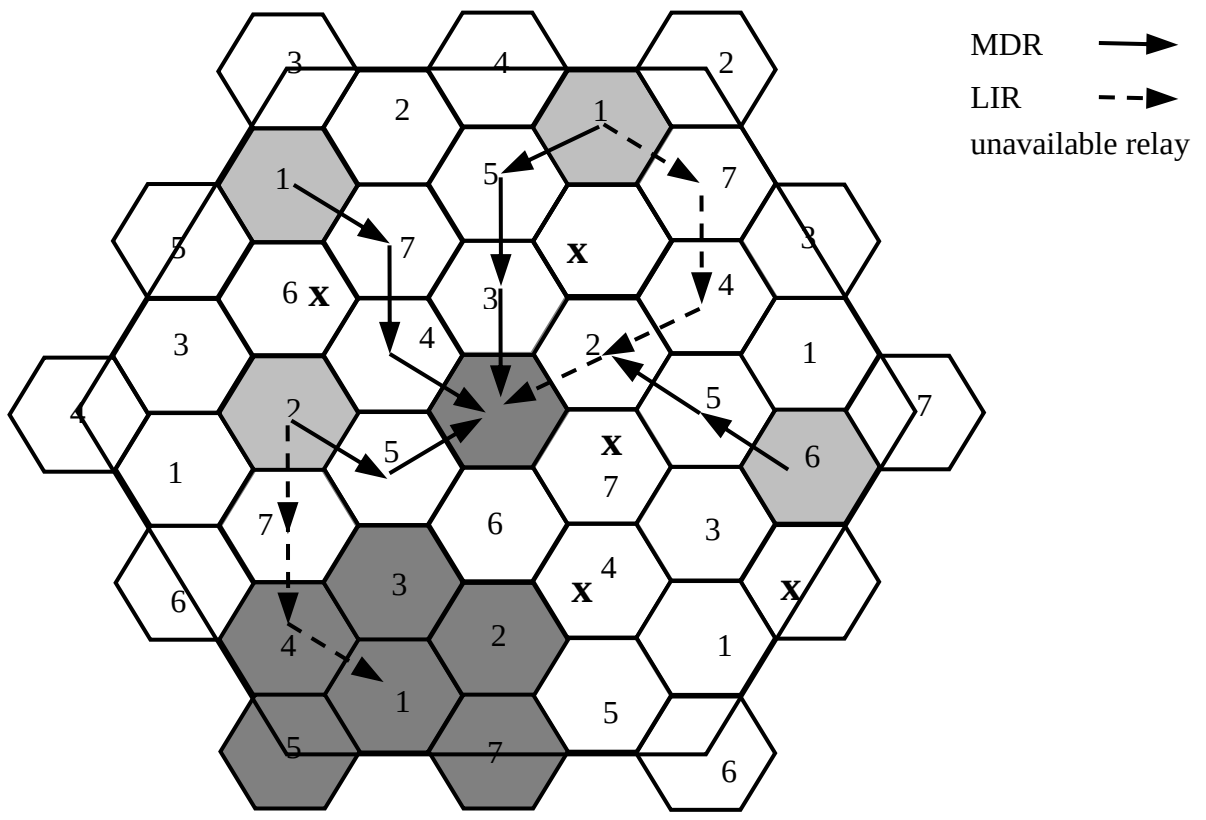

Fig. 3. Routing/scheduling for $m^3$ network by clustering factor $K$=7.

We let the users of the same type to share the slot. The transmission turn (in a round robin fashion) is given by the index of user type. The distance between the simultaneously transmitting and receiving nodes and so, the interference level is given by the network topology. For a given $R$, this parameter can be absorbed again by parameter $H$. This will be referred to as scheduling state *2* denoted as $ss^{(2)}$ and the overall scheduling interval of 7 time slots as $T^{(2)} = 7$. The drawback of this scheme is that there may be slots when there is only one transmission or very few transmissions. To eliminate this drawback, a new routing/scheduling protocol is suggested in the next subsection.

### *B. Limited Interference Routing/Scheduling (LIR)*

By considering the clustering scheme shown in Fig.3, LIR protocol relies on the fact that highest interference distance (minimum interference level) is obtained when the slot is shared between users of the same type $k$, $k = 1,\dots, K$. This parameter can be again absorbed in the model by $H$. So, whenever is possible, the users relay to the adjacent user from the type that is simultaneously available to all users since they are located on the largest possible distance. This will be referred to as scheduling state 1 denoted as $ss^{(1)}$ and the overall scheduling interval of 1 time slots as $T^{(1)} = 1$. An example of this routing protocol is shown in Fig. 3 where the limited interference routes are indicated with dashed lines. We can see that the users relay to adjacent users of type $k$=7 and then of type $k$=4 whenever is possible. In the case when the adjacent relay from the same user type is located at ring $h' > h$, the user will not choose this option in order to avoid the loop in the route. This is the case of the transmitter of type $k$=6 (light shadowed subcell in Fig.3). So, in this protocol only one time slot $(T^{(1)} = 1)$ is needed for simultaneous one hop transmission on all routes, as opposed to $K$=7 slots used in the round robin scheduling in MDR protocol, i.e. $T^{(2)} = KT^{(1)}$.

The relaying alternatives when using LIR protocol are shown in Fig.4. First, the protocol tries to operate in $ss^{(1)}$ mode. This requires that all $N/K$ users of the same type are available at the same time. This occurs with probability $p^{N/K}$ as shown in Fig. 4 (right hand side) where $N$ is the number of subcells and $K$ is the tessellation factor. If available, relaying will take place as indicated in Fig. 4 for state $ss^{(1)}$. If this option is not available, which occurs with probability $p_0^{(1)}$, the protocol will switch to operate in state $ss^{(2)}$ as indicated in the same figure (left hand side). The $ss^{(2)}$ follows MDR protocol as described in the previous subsection. For different hops on the routes, the protocol may alternate between the states $ss^{(1)}$ and $ss^{(2)}$. The relaying subcell transmission probabilities for initial states $ss^{(1)}$ and $ss^{(2)}$ are presented in Figs. 5 and 6, respectively. In Fig. 5, the procotol will remain in state $ss^{(1)}$ with probability $p_n^{(1)}(1-p_0^{(1)})$, $n$=1,…,6 where $p_n^{(1)}$ is obtained as $p_n^{(1)} = p^{N/K}(1-p^{N/K})^{n-1}$ and $p_0^{(1)} = 1 - \sum_n p_n^{(1)}$. Otherwise, the protocol will move to state $ss^{(2)}$ with probability $p_n^{(1)}p_0^{(1)}$, $n$= 1,…,6. In Fig. 6, when the initial state is $ss^{(2)}$, the protocol will remain in state $ss^{(2)}$ with probability $p_n^{(2)}p_0^{(1)}$, $n$= 1,…,6 where $p_n^{(2)} = p(1-p)^{n-1}$, and it will move to state $ss^{(1)}$ with probability $p_n^{(2)}(1-p_0^{(1)})$, $n$=1,…,6.

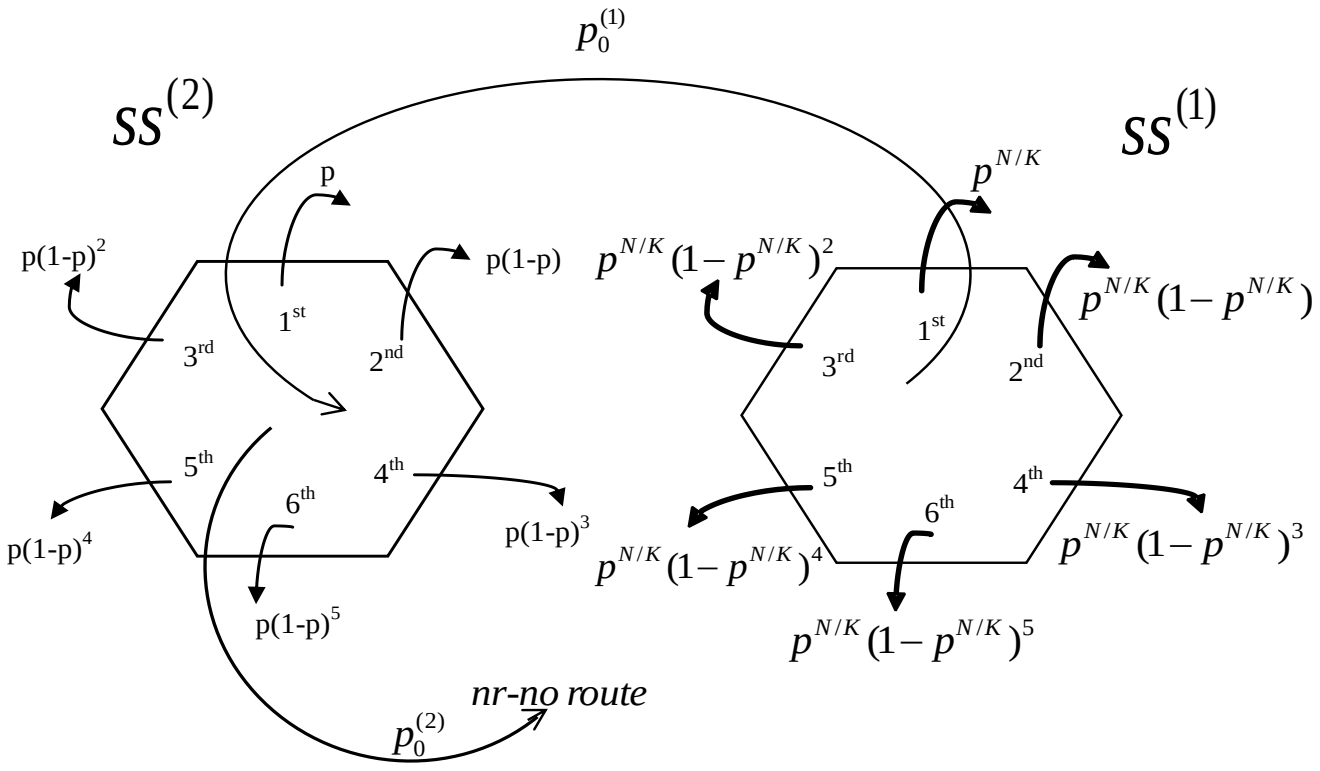

Fig. 4. Modeling Limited Interference Routing/Scheduling (LIR).

## V. ANALYSIS OF $m^3$ ROUTE DISCOVERY PROTOCOLS BY COMPRESSED NETWORK MODEL

For the analysis of the route discovery protocols, we map the tessellation scheme into an absorbing Markov chain, where the BS/AP denotes the absorbing states.

In general relaying from subcell $i$ to subcell $j$ will take place with probability $p_{ij}$ which can be arranged in a subcell relaying probability matrix $\mathbf{P} = \|p_{ij}\| = \|p(h,\theta;h',\theta')\|$ where the first set of indexes ($h$,$\theta$) refers to the location of the transmitter and the second one ($h'$,$\theta'$) to the location of the receiver. The mapping $i \rightarrow (h,\theta)$ and $j \rightarrow (h',\theta')$ is illustrated in Fig.7.

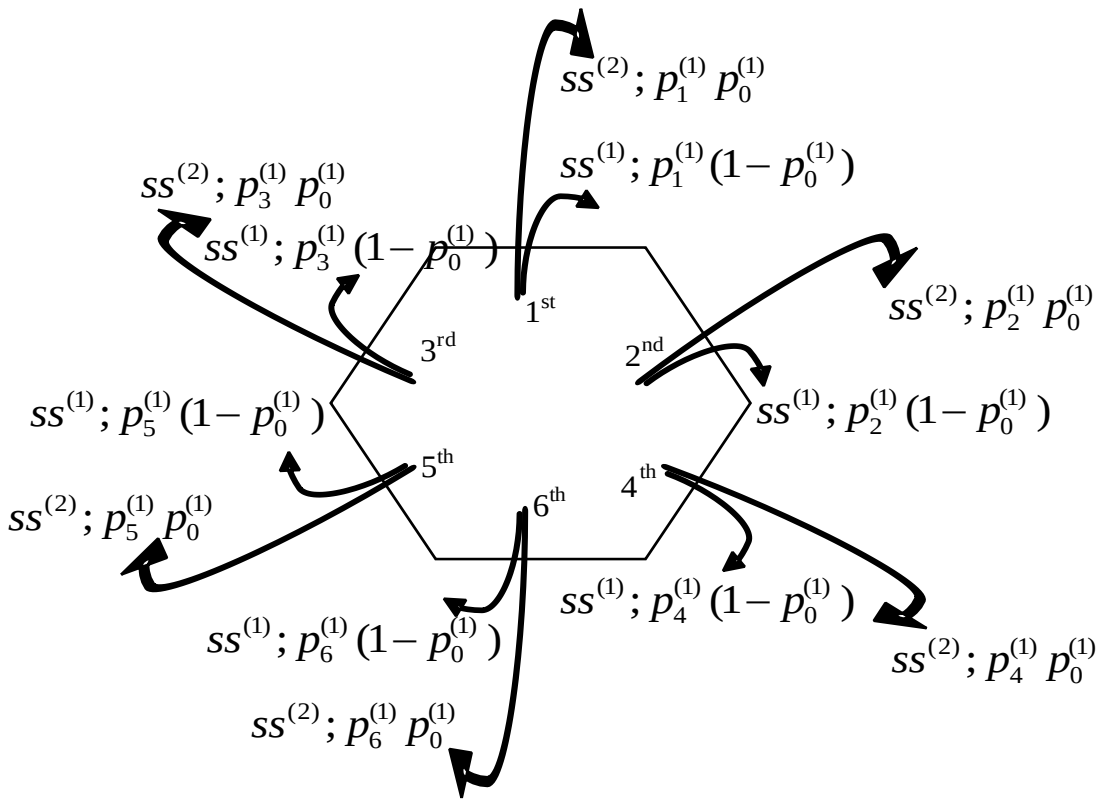

$$p_n^{(1)} = p^{N/K}(1-p^{N/K})^{n-1}, n = 1,...,6$$

Fig. 5. Relaying transmission probabilities for initial state $ss^{(1)}$.

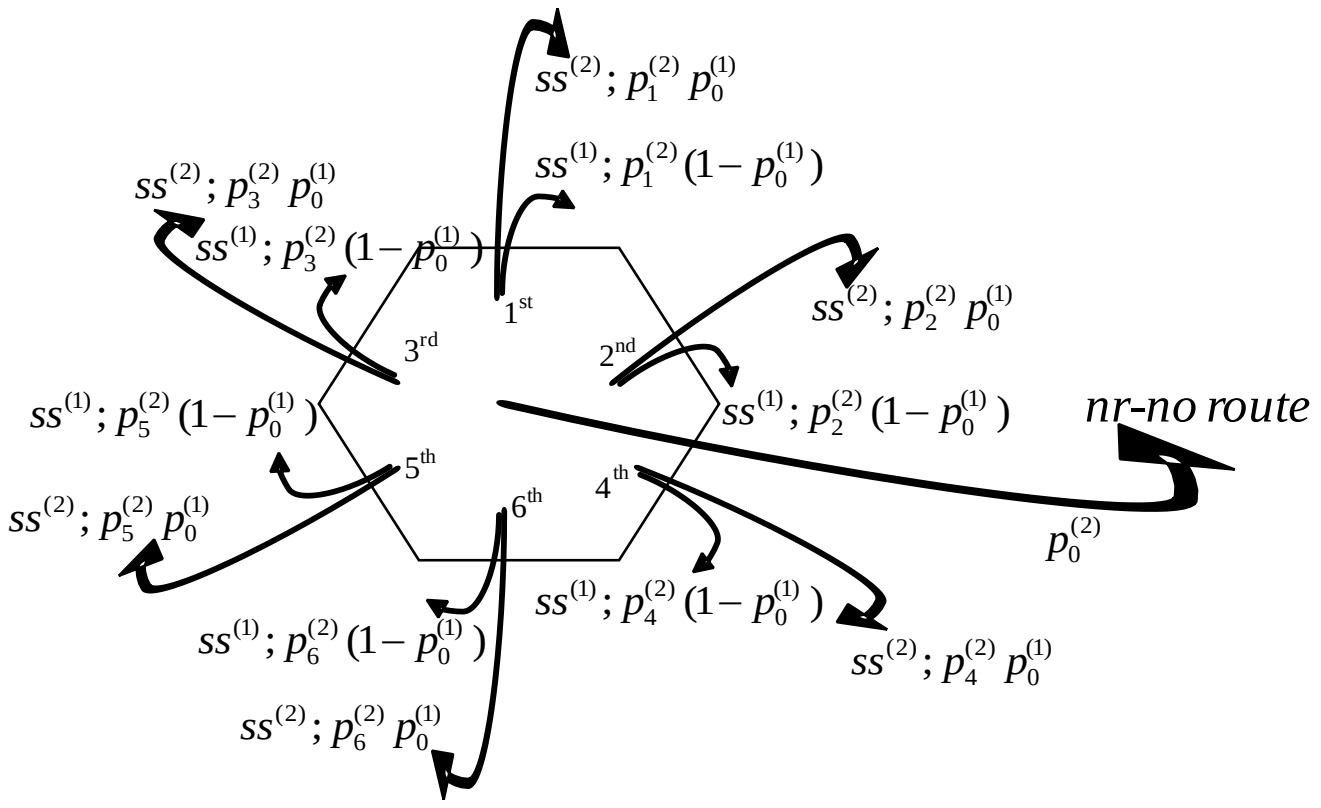

$$p_n^{(2)} = p(1-p)^{n-1},\ n = 1,...,6;\ \ T^{(2)} = KT^{(1)}$$

Fig. 6. Relaying transmission probabilities for initial state $ss^{(2)}$.

Following the MDR scheme presented in Fig. 2, in the sequel we derive general expressions for the subcell transition probabilities under the assumption that the scheduling protocol imposes constant dwell time in each subcell. These expressions can be obtained for other transmissions priorities, *i.e.* LIR protocol, by using the same reasoning.

The entries of the relaying probability matrix **P** are obtained as $p(h,\theta;h',\theta') = p_n = p(1-p)^{n-1}$ where $p$ is obtained by the aggregation process in compressed state vector $\boldsymbol{\nu}^{(c)}$, $h \le h' \le H$ and $n$=1,…,6. Thus, the overall relaying probability to any adjacent subcell is $p_t = \sum_n p_n$. The probability that the user does not relay to any other user is denoted by $p_0$, $p_0 = 1 - p_t$ which is transferred to an additional absorbing state $nr$ (no route).

Then, we reorganize the relaying probability matrix into a $(N+1)\times(N+1)$ matrix of the form [26]

$$\mathbf{P}^* = \begin{bmatrix} \mathbf{I} & \mathbf{0} \\ \mathbf{R} & \mathbf{Q} \end{bmatrix} \quad (5)$$

where $N$ is the number of subcells, **I** is ($N_A$+1)x($N_A$+1) diagonal unitary matrix corresponding to the number of absorbing states including $N_A$ BS/APs plus no route state $nr$, **0** is ($N_A$+1) x ($N$-$N_A$) all zero matrix, **R** is ($N$-$N_A$) x ($N_A$+1) matrix of transition probabilities from transient states to absorbing states and **Q** is ($N$-$N_A$) x ($N$-$N_A$) matrix of transition probabilities between transient states. By using notation $\mathbf{N}=(\mathbf{I}-\mathbf{Q})^{-1}$, the mean time for the process to reach any absorbing state (BS/AP or $nr$) starting from transient state $i$ (subcell $i$) is [26]

$$\boldsymbol{\tau}=(\tau_0,\tau_1,...,\tau_{N-N_A-1})^{tr}=T(\mathbf{I}-\mathbf{Q})^{-1}\mathbf{1}=T\mathbf{N1} \qquad (6)$$

when the dwell time $T_i$ for each state $i$ is the same, $T_i = T$, and $(\cdot)^{tr}$ denotes the transpose operation. Otherwise, $\boldsymbol{\tau}=(\tau_0,\tau_1,...,\tau_{N-N_A-1})^{tr}=(\mathbf{I}-\mathbf{Q})^{-1}\mathbf{e}=\mathbf{Ne}$ where $\mathbf{e}=column\,vec\{T_i\}$ and **1** is $(N-N_A)\times 1$ column vector of all ones. For the normalized dwell time $T_i = T$=1, the entrees $\tau_i$ of vector $\boldsymbol{\tau}$ represent the average number of hops from state $i$ (subcell $i$) to absorbing state (BS/AP or $nr$). This expression is used in Section VI to obtain the transmission delay in the definition of the utility function. In general, the variance of that time is

$$\mathrm{var}\,\boldsymbol{\tau}=2(\mathbf{I}\text{-}\mathbf{Q})^{-1}\mathbf{TQ}(\mathbf{I}\text{-}\mathbf{Q})^{-1}\mathbf{e}+(\mathbf{I}\text{-}\mathbf{Q})^{-1}(\mathbf{e}_{sq})-((\mathbf{I}\text{-}\mathbf{Q})^{-1}\mathbf{e})_{sq} \qquad (7)$$

where $\mathbf{T}=diag\,matrix\{T_i\}$, and if the dwell times are the same

$$\mathrm{var}\,\boldsymbol{\tau}=[(2\mathbf{N}\text{-}\mathbf{I})\mathbf{N1}-(\mathbf{N1})_{sq}]T^2 \qquad (8)$$

where $(\mathbf{N1})_{sq}$ is the square of each component of **N1**. The average time to reach an absorbing state is

$$\tau_a=\mathbf{f}\boldsymbol{\tau} \qquad (9)$$

where **f** is a row vector of probabilities of users' initial positions and $\boldsymbol{\tau}$ is a column vector given by (6). The probability that the Markov process starting in transient state $i$ ends up in absorbing state $j$ is $b_{ij}$, and it is obtained as [26]

$$\mathbf{B}=\left[b_{ij}\right]=(\mathbf{I}-\mathbf{Q})^{-1}\mathbf{R} \qquad (10)$$

The average probabilities of hand off, accessing the base station and no route are given as

$$\mathbf{p_{ac}}=(p_{wlan},p_{bs},p_{nr})=\mathbf{fB} \qquad (11)$$

where **f** is the vector of probabilities of initial user positions.

In the case of LIR protocol, the analysis remains the same except that the number of states in the absorbing Markov chain is doubled since each subcell can be either in $ss^{(1)}$ or $ss^{(2)}$ state.

The complexity of the protocols, in terms of number of iterations needed to find the route for a given user to the AP, can be obtained by including a new separate state in the Markov model.

## VI. Traffic Offloading Incentives

Once the available routes, for a given $H$, are found as discussed in Section IV, we measure the performance of the network in terms of the utility function that includes a number of details specific for this network. The optimum tessellation/topology is obtained by using the compressed control mechanism presented in Section III. Then, a cooperative multi-operator call/session access policy is developed. The policy is based on *dynamic micro-economic criteria* for cooperation decision in the $m^3$ *network.* The interest in traffic offloading from the cellular network to a local WLAN is quantified by the offset in the network utility function before and after the hand offs of certain number of users from the cellular network into the WLAN (or small cell). In this segment we assume two different operators. The cellular network operator will be referred to as mobile network operator (MNO) and the WLAN operator as small service operator (SSO). The offloading price, used as basis for access decision, is dynamically changed based on the instantaneous number of new/ended calls in the cell and WLAN, and offloaded calls to the WLAN. In this way, the terminating sessions in both cellular network and WLAN are also incorporated into the overall model of the system. These factors have impact on the offloading price due to the change of the interference/capacity, delay and power consumption in the network. To reflect these effects, the utility function for the MNO before offloading will include:

- The capacity of user $i$ on the route $\Re_i$ towards the BS. This is given by

$$C_{\Re_i}=\min c_l\left(\pi_i\right),\ l\in\Re_i \qquad (12)$$

where $c_l$ is obtained as in Section II, for a given scheduling $\pi_i \in\mathbf{\Pi}_{bs}$; $\mathbf{\Pi}_{bs}$ is the set of feasible scheduling at the BS.

- The transmission delay $D_{\Re_i}\left(\pi_i\right)$ for user $i$ to transmit the packet on route $\Re_i$. When MDR protocol is used, $D_{\Re_i}=K\tau_i$ where $K$ is the tessellation factor and $\tau_i$ is obtained by (6) for normalized dwell time $T$=1. Instead, if LIR protocol is used, the delay may be reduced, as already explained in Section IV, so

$$D_{\Re_i}\in[\tau_i,K\tau_i]. \qquad (13)$$

It is worth noticing that the cooperation between different operators has impact on the delay through parameter $\tau_i$, which is obtained based on the aggregated relaying probability $p$.

- The path cost which reflects the overall power consumption on route $\Re_i$ of effective length $h_{ei}$,

$$cost_{\Re_i}=Ph_{ei} \qquad (14)$$

If we assume that the dwell time is constant for each subcell, $T_i$=$T$=1, then $h_{ei}$ is equal to the normalized mean time $\tau_i$ for user $i$ to access the BS as defined by (6).

Then, the utility for the MNO before offloading can be written as

$$U=\sum\nolimits_{i\in\mathrm{N}_{bs}}U_i,\quad U_i=\rho C_{\Re_i}/\left(D_{\Re_i}cost_{\Re_i}\right),\Re_i\in\Re_{bs} \qquad (15)$$

where $\mathrm{N}_{bs}$ is the set of users in the cellular network, $\Re_{bs}$ is the set of routes towards the BS and $\rho$ is the revenue per unit of the utility function. In general the maximization of the utility function in (15) will drive all parameters in the right direction: large capacity, small delay and small cost, usually representing the power consumption. For the network guaranteeing quality of service (QoS) further specifications in the form of constraints might be needed like $C_{\Re_i} \geq C_0$ *and / or* $D_{\Re_i} \leq D_0$ *and / or even* $cost_{\Re_i} \leq cost_0$. The relaxations of these constraints (like Lagrangian method) will modify the utility function accordingly and solution of such problems are elaborated in the literature [27]. Further modifications of the utility function (15) are also obtained if the revenue $\rho$ is proportional to the quality of service (proportional to the capacity and inversely proportional to delay and cost). With all these options in mind in the sequel we still use relatively simple form of the utility function and focus on the compressed optimization which is valid for any form of the utility.

The routing schemes defined in Section IV include also some heuristics for the scheduling. Otherwise, in order to control the interlink interference we have to optimize the subset of simultaneously active links. As already mentioned, the optimization of the scheduling in multihop networks is NP hard. So, in the sequel we will adopt the scheduling heuristics presented in Section IV which allows us to use the utility defined as in (15) and further specify (16) as presented below.

### *A) Tessellation/Topology Optimization*

Similar to Table II, the compressed network state vector $\mathbf{v}^{(c)}(H, P, p) \in \mathbb{V}^{(c)}$ is considered. By using the utility function defined in (15), we can simultaneously optimize the system throughput, power consumption and delay, as a function of the number of rings *H* and transmission power *P*. The optimum tessellation/topology is obtained by solving the following optimization problem,

$$\begin{aligned} &\underset{P,H}{\text{maximize}} \quad U = \sum_{i \in N_{bs}} U_i = \rho C_{\Re_i} / \left( D_{\Re_i} cost_{\Re_i} \right) \\ &\text{subject to} \quad \Re_i \in \Re_{bs} \qquad (16) \\ &\qquad C_{\Re_i} = C_{\Re_i}(\pi_i), D_{\Re_i} = D_{\Re_i}(\pi_i), \pi_i \in \mathbf{\Pi}_{bs} \end{aligned}$$

where the capacity $C_{\Re_i}$ and path cost $cost_{\Re_i}$ are given by (12) and (14), respectively, and $D_{\Re_i} = K\tau_i$ with $\tau_i$ obtained by (6) for normalized dwell time *T*=1. The route $\Re_i$ towards the BS belongs to the feasible set of routes in the cell $\Re_{bs}$. The capacity $C_{\Re_i}$ and $D_{\Re_i}$ are constrained by the scheduling set $\mathbf{\Pi}_{bs}$. The optimum tessellation will be used in the following sections to optimize traffic offloading.

### *B) MNO Incentives*

The utility for the mobile network operator before offloading is given by (15). After offloading, the utility is denoted by $U'$, and is formally defined as

$$U' = \sum_{i' \in \mathsf{N}'_{bs}} U_{i'}, \qquad (17)$$

where $\mathsf{N}'_{bs} = \mathsf{N}^+_{bs} \cup \mathsf{N}_\mu$, $\mathsf{N}^+_{bs}$ is the set of users transmitting to the BS in the next instant (after offloading) defined by (3a). The utility per user *i'* after offloading is

$$U_{i'} = \begin{cases} \rho C_{\Re_{i'}} / \left( D_{\Re_{i'}} cost_{\Re_{i'}} \right), \Re_{i'} \in \Re^+_{bs}, & if\ i' \in \mathsf{N}^+_{bs} \\ (\rho - \chi) C_{\Re_{1i'}} / \left( D_{\Re_{1i'}} cost_{\Re_{1i'}} \right), \Re_{1i'} \in \Re_\mu, & if\ i' \in \mathsf{N}_\mu \end{cases}$$

From the above definition of the utility it is worth noting that,

- For those users that remain in the cell after the offloading decision ($i' \in \mathsf{N}^+_{bs}$), their utility is defined as before offloading (15) but the value obtained will be different as the traffic in the network has changed. The new route $\Re_{i'}$ belongs to the set of routes in the cellular network after offloading $\Re^+_{bs}$ and, the capacity $C_{\Re_{i'}}$, delay $D_{\Re_{i'}}$ and path cost $cost_{\Re_i}$ are given by (12)-(14).
- For those users that have been offloaded ($i' \in \mathsf{N}_\mu$), the revenue obtained by the MNO, $\rho$, is now decreased by the price paid to the SSO for offloading, $\chi$. The route for the offloaded user $\Re_{1i'}$ belongs to the feasible set of offloaded routes $\Re_\mu$. It is worth noting that the route $\Re_{1i'}$ may consist of links from the macrocell and WLAN. The capacity $C_{\Re_{1i'}}$, delay $D_{\Re_{1i'}}$ and path cost are obtained again by (12)-(14), respectively, where the number of hops towards the WLAN is $m_{e1i'}$.

The aim of the MNO is to maximize the offset of the utility function, after and before handoff, for the offloading price $\chi$ offered by the SSO as

$$\begin{aligned} &\underset{\mathsf{N}_\mu}{\text{maximize}} \quad \Delta U = U' - U = \sum_{i' \in \mathsf{N}'_{bs}} U_{i'} - \sum_{i \in \mathsf{N}_{bs}} U_i \\ &\qquad = \sum_{i' \in \{\mathsf{N}^+_{bs} \cup \mathsf{N}_\mu\}} U_{i'} - \sum_{i \in \mathsf{N}_{bs}} U_i \\ &\text{subject to} \quad \mathsf{N}_\mu \subset \mathsf{N}_{bs} \qquad (18) \\ &\qquad \mathsf{N}^+_{bs} = \mathsf{N}_{bs} \cup \mathsf{N}_{\lambda bs} \setminus \mathsf{N}_{\mu bs} \setminus \mathsf{N}_\mu \\ &\qquad \pi_i, \pi_{i'} \in \mathbf{\Pi}_{bs};\ \pi_{1i'} \in \mathbf{\Pi}_{wlan} \end{aligned}$$

with respect to the set of offloaded users $\mathsf{N}_\mu$. The scheduling sets at the BS, $\mathbf{\Pi}_{bs}$, and WLAN, $\mathbf{\Pi}_{wlan}$, include the scheduling options provided by MDR or LIR protocols. So, the optimization problem is solved by using any of these routing and scheduling heuristics and evaluating the utility function for the possible routes until the maximum utility is obtained.

We assume that the optimization problem described by (18) is solved for a given offloading price $\chi$ that the SSO will provide to the MNO in the negotiation process. This process is elaborated in details in Section VI.C. The MNO obtains the optimum set of users to offload through the SSO, $\mathsf{N}^*_\mu$, at a given price $\chi$ which is affected by the current and new users arrivals to the cell $\mathrm{N}_{\lambda bs}$. The offset in the utility also depends on the position of the users in the network.

*C) SSO Incentives*

We assume the same network architecture for the SSO. The utility for SSO before offloading is denoted by $U_1$ and defined as

$$U_1 = \sum_{i\in_{wlan}} U_{1i}, \quad U_{1i} = \rho_1 C_{\Re_{1i}} / \left(D_{\Re_{1i}} cost_{\Re_{1i}}\right), \Re_{1i} \in \Re_{wlan} \quad (19)$$

where $\mathrm{N}_{wlan}$ and $\Re_{wlan}$ are the set of users and routes in the WLAN, respectively and $\rho_1$ is the revenue per unit of the utility function in the WLAN. The capacity $C_{\Re_{1i}}$, delay $D_{\Re_{1i}}$ and path cost $cost_{\Re_{1i}}$ are obtained as in (12)-(14) for route $\Re_{1i}$ towards the WLAN of effective length $m_{e1i}$.

After the handoff, the utility for the SSO is given by

$$U_1' = \sum_{i'\in \mathrm{N}_{wlan}^{+}} U_{1i'} \quad (20)$$

where $\mathrm{N}_{wlan}^{+}$ is the set of users in the WLAN in the next instant (after offloading) defined by (3). For each particular user *i'*, the utility is obtained as

$$U_{1i'} = \begin{cases} \rho_1 C_{\Re_{1i'}} / \left(D_{\Re_{1i'}} cost_{\Re_{1i'}}\right), & \Re_{1i'} \in \Re_{wlan}^{+} \setminus \Re_{\mu};\ if\ i' \in \mathrm{N}_{wlan}^{+} \setminus \mathrm{N}_{\mu} \\ \chi C_{\Re_{1i'}} / \left(D_{\Re_{1i'}} cost_{\Re_{1i'}}\right), & \Re_{1i'} \in \Re_{\mu}; if\ i' \in \mathrm{N}_{\mu} \end{cases}$$

- For those users that were already in the WLAN before the offloading decision ( $i' \in \mathrm{N}_{wlan}^{+} \setminus \mathrm{N}_{\mu}$), their utility is defined as before offloading (19) but the value obtained will be different due to the traffic changes in the network. The new route $\Re_{1i'}$ belongs to the set of routes in the WLAN after offloading $\Re_{wlan}^{+} \setminus \Re_{\mu}$ and, $C_{\Re_{1i'}}$, $D_{\Re_{1i'}}$ and $cost_{\Re_{1i'}}$ are given by (12)-(14) where number of hops towards the WLAN is $m_{e1i'}$.
- For those users that have been offloaded ($i' \in \mathrm{N}_{\mu}$), the price charged by the SSO for offloading is given by $\chi$. The route for the offloaded user $\Re_{1i'}$ belongs to the feasible set of offloaded routes $\Re_{\mu}$. The rest of the parameters are obtained as before.

The aim of the SSO is to maximize the offset of the utility function after and before handoff as

$$\underset{\chi,\mathrm{N}_{\mu}}{\text{maximize}} \quad \Delta U_1 = U_1' - U_1 = \sum_{i'\in \mathrm{N}_{wlan}^{+}} U_{1i'} - \sum_{i\in \mathrm{N}_{wlan}} U_{1i}$$

$$\text{subject to} \quad \mathrm{N}_{wlan}^{+} = \mathrm{N}_{wlan} \cup \mathrm{N}_{\lambda wlan} \setminus \mathrm{N}_{\mu wlan} \cup \mathrm{N}_{\mu} \quad (21)$$

$$\pi_{1i}, \pi_{1i'} \in \mathbf{\Pi}_{wlan}$$

$$\rho_1 \le \chi \le \rho$$

with respect to the cost of handoff per user $\chi$ and the set of offloaded users, $\mathrm{N}_{\mu}$. The capacity on the route towards the WLAN, before and after handoff, are given by $C_{\Re_{1i}}$ and $C_{\Re_{1i'}}$, respectively and are constrained by the scheduling set $\mathbf{\Pi}_{wlan}$. The same applies for the delay on those routes $D_{\Re_{1i}}$ and $D_{\Re_{1i'}}$. The path cost, before and after handoff, $cost_{\Re_{1i}}$ and $cost_{\Re_{1i'}}$, respectively depends on the power consumption on the path and route length towards the WLAN, $m_{e1i}$ and $m_{e1i'}$. The offloading cost $\chi$ should be lower/equal than the revenue received per user at the MNO and larger/equal than the revenue at the SSO.

For the set of users $\mathrm{N}_{\mu}$ that the MNO has decided to offload, the SSO solves the optimization problem (21) to obtain the optimum price $\chi$ which is affected it by the current and new users arriving at the WLAN $\mathrm{N}_{\lambda wlan}$. Again these parameters depend on the location of the users. The optimization problem is solved as before for MDR or LIR protocols.

*D) Collaborative negotiation between MNO and SSO*

The negotiation process between MNO and SSO to choose the offloading price $\chi$ is described in the following steps:

1. SSO offers the price for the service $\chi$
2. MNO calculates $\Delta U(\chi, \mathrm{N}_{\mu})$ by (18) and pass it to SSO
3. SSO calculates $\Delta U_1(\chi, \mathrm{N}_{\mu})$ by (21) and offers new price $\chi'$ based on the following relation between $\Delta U$ and $\Delta U_1$:

$$\chi' = \begin{cases} \chi - \Delta\chi;\ \Delta U_1 > \Delta U \\ \chi + \Delta\chi;\ \Delta U_1 < \Delta U \end{cases}$$

$$\chi = \chi'$$

4. The process iterates until $\Delta U(\chi, \mathrm{N}_{\mu}) = \Delta U_1(\chi, \mathrm{N}_{\mu})$ and then, the optimum price is obtained $\chi = \chi^*$.

Another option is to change simultaneously $\mathrm{N}_{\mu}$ and $\chi$ as:

1. SSO offers the price for the service $\chi$
2. MNO calculates $\Delta U(\chi, \mathrm{N}_{\mu})$ by (18) and pass it to SSO
3. SSO calculates $\Delta U_1(\chi, \mathrm{N}_{\mu})$ by (21) and offers new $\chi'$

$$\chi' = \begin{cases} \chi - \Delta\chi;\ \Delta U_1 > \Delta U \\ \chi + \Delta\chi;\ \Delta U_1 < \Delta U \end{cases}$$

$$\chi = \chi'$$

4. MNO calculates $\Delta U(\chi, \mathrm{N}_{\mu})$ and offers new $\mathrm{N}_{\mu}'$

$$\mathrm{N}_{\mu}' = \begin{cases} \mathrm{N}_{\mu} \setminus \Delta\mathrm{N}_{\mu};\ \Delta U_1 < \Delta U \\ \mathrm{N}_{\mu} \cup \Delta\mathrm{N}_{\mu};\ \Delta U_1 > \Delta U \end{cases}$$

$$\mathrm{N}_{\mu} = \mathrm{N}_{\mu}'$$

5. Process iterates until $\Delta U(\chi, \mathrm{N}_{\mu}) = \Delta U_1(\chi, \mathrm{N}_{\mu})$ and then, the optimum price is obtained $\chi = \chi^*$.

The process can be further extended to include possible variations in the set $\mathrm{N}_{\lambda bs}$ and $\mathrm{N}_{\lambda wlan}$ representing the number of newly accepted sessions in the BS and WLAN, respectively. Some illustrations of this possibility will be provided in next section.

## VII. Performance Evaluation

We present some numerical results to evaluate the performance of the $m^3$ route discovery protocols and the cooperative multi-operator call/session access policies based on the proposed economic models. Single technology and multi-technology scenarios are considered with different sets of available users. The results are obtained in Matlab for a macrocell of radious $R$ =1000 $m$ and $K$ = 7. The path loss exponent $\alpha$ = 2 and the noise power is $N_r = 10^{-4}$ W/Hz.

As a first step in Fig. 8, the utility function defined as in (16) is presented versus $H$. By using model compression, we can see that the feasible choices are $H$ = 5, $P$ = 0.1; $H$ = 4, $P$ = 0.15-0.25 and $H$ = 3, $P$ = 0.3-0.4. Further, we assume that a user $i$ can successfully transmit to its adjacent relay $j$ when the received power at $j$ exceeds the receiver sensitivity $\varepsilon$ (depending on the noise level). For a given relaying distance $d_r$, the minimum transmission power for user $i$ is $P_{i,min} = \varepsilon\cdot(d_r)^\alpha$, where $\alpha$ is the path loss factor. Users are interested in transmitting with the minimum power possible $P_i = P_{i,min}$ to reduce interference and power consumption. For simplicity, we assume that the tessellation factor $r$ is the same for all subcells and thus, users transmit with the same power $P_i = P$. For these reasons, unless otherwise stated, the simulations are based on the scenario presented in Fig. 1 for the optimum tessellation given by $H$ = 4 and $P$ = 0.15.

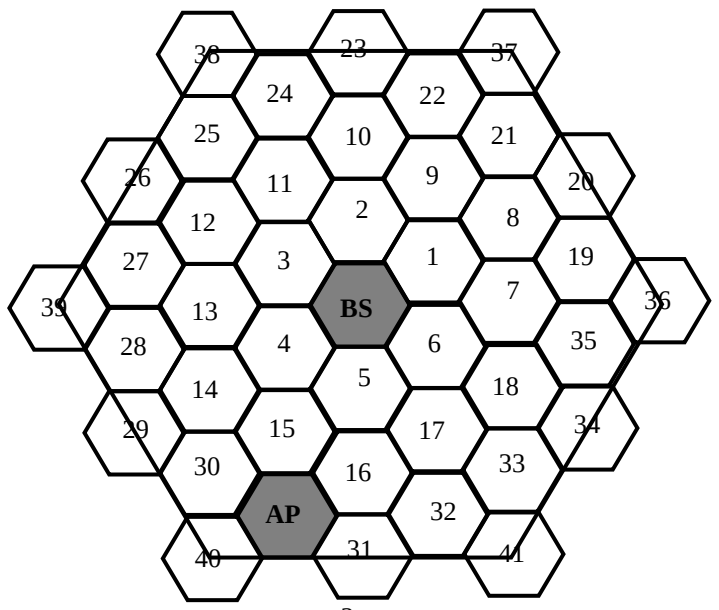

Fig. 7. $m^3$ scenario.

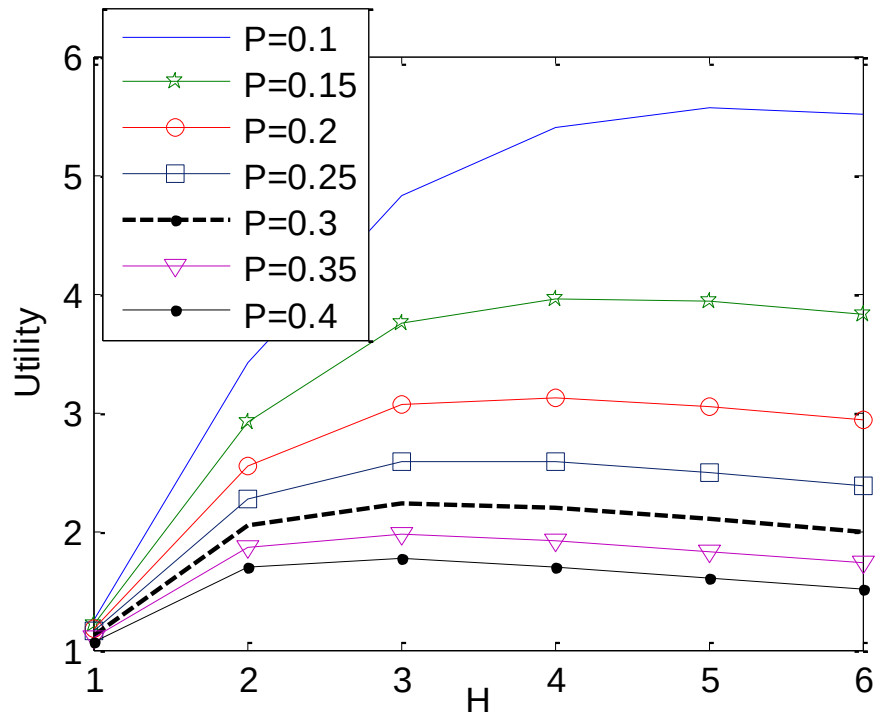

Fig. 8. Utility versus $H$ for different power values.

### *A) $m^3$ route discovery protocols*

The scenario considered is presented in Fig. 7 where the WLAN AP is located in subcell $h$ = 3, $\theta$ = 250. In Fig. 9 and 10, the average message delivery time $\tau_i$ is presented versus the subcell index $i$ for MDR and LIR, protocols, respectively. The subcell index $i$ corresponds to subcell number in the multi-technology scenario shown in Fig. 7. The users from index $i$ = 1 to 6, are located in ring with index $h$ = 1, users from $i$ = 7 to 18 in $h$ = 2, and so on. The oscillations in the results within the same hop are due to the hexagonal tessellation which indicates that the distance on a chosen route from the users to the BS in the same hop may change. We assume that the dwell time for MDR protocol is $T = K = 7$ and for LIR protocol, $T = 1$ . So, we can see that when $p$=1, $\tau_{\mathrm{MDR}}$ is 7 times larger than $\tau_{\mathrm{LIR}}$. For other values of $p$<1, $\tau_{\mathrm{MDR}}$ is approximately 2.5 times larger than $\tau_{\mathrm{LIR}}$. As before, $\tau_i$ significantly decreases for those users closer to the WLAN.

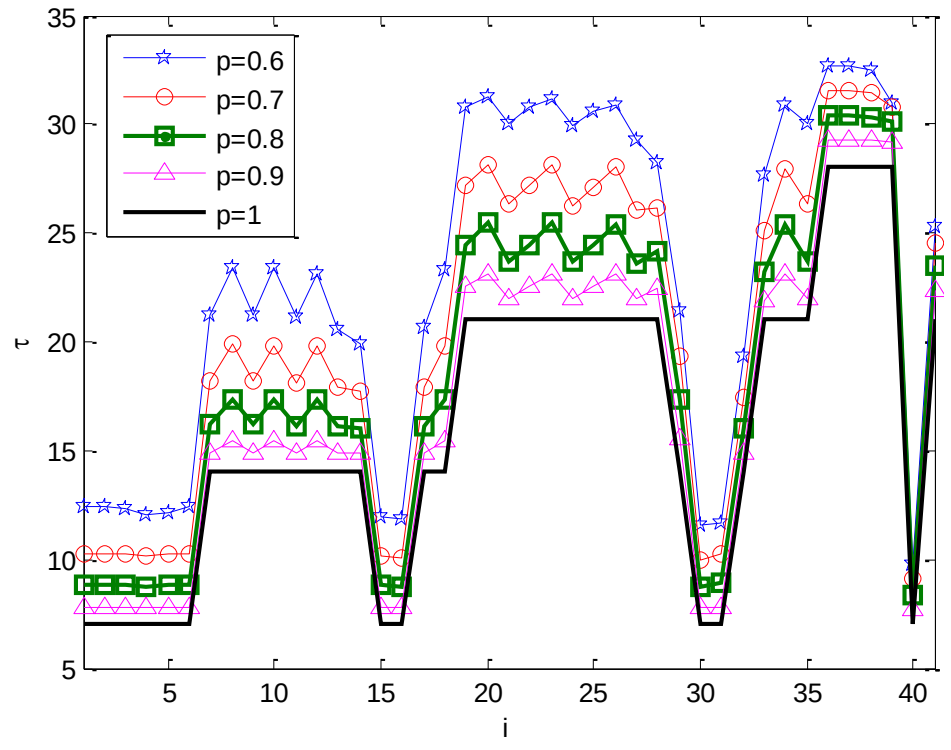

Fig. 9. $\tau_i$ vs. subcell index $i$ for MDR.

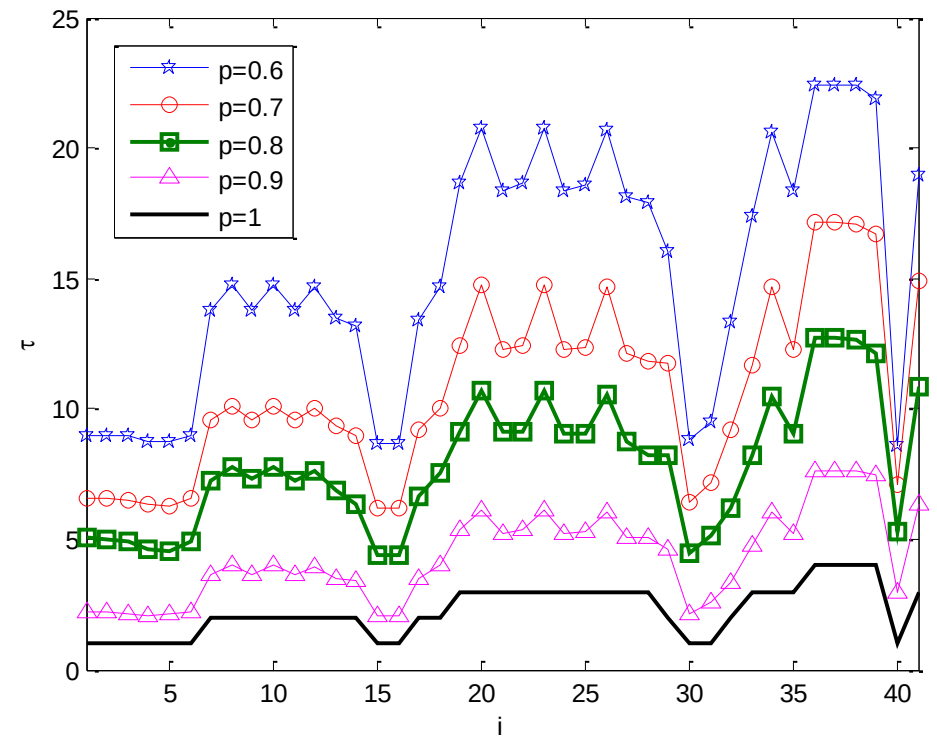

Fig. 10. $\tau_i$ vs. subcell index $i$ for LIR.

In Fig. 11, the probability $B$ of selecting the BS/AP, which is obtained by (10), is presented together with the probability of no route for the same scenario by using MDR protocol. We can see that the probability that the users reach the BS, $B_{\mathrm{BS}}$, decreases for the users closer to the WLAN. For those users, $B_{\mathrm{WLAN}} > B_{\mathrm{BS}}$. The opposite behavior is observed for the users closer to the BS. The probability of no route, $B_{nr}$, increases for the users located far from any BS or AP.

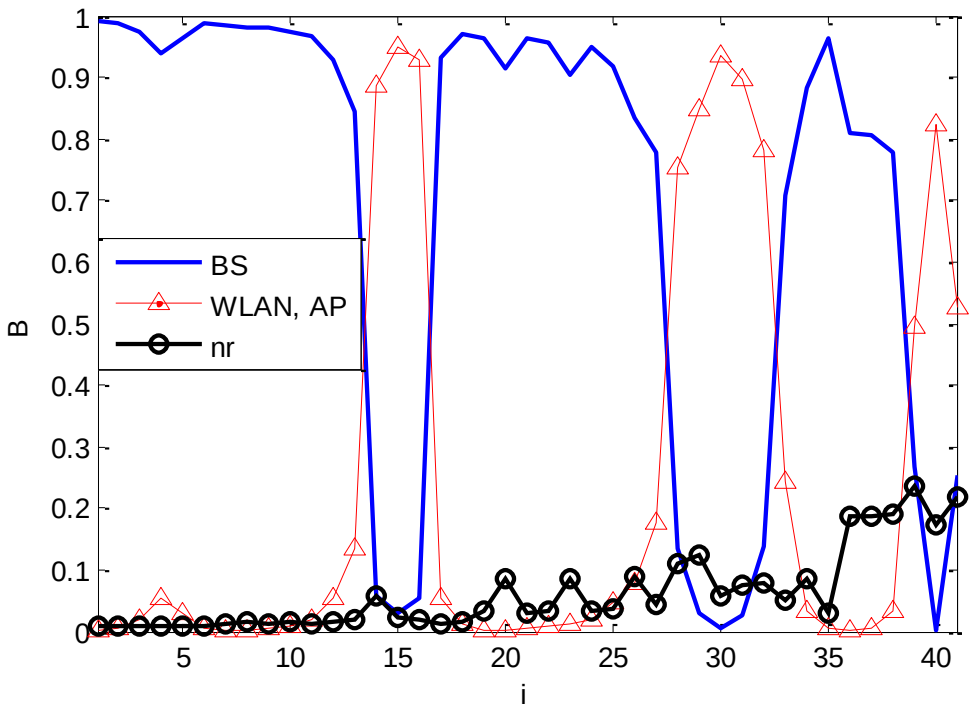

Fig. 11. $B_i$ vs. the subcell index $i$ for the scenario shown in Fig. 7.

### *B) Capacity and throughput for the modified m*$^3$ *route discovery protocols*

As already discussed, MDR protocol has the advantage that the number of iterations needed by the protocol to find the route is significantly lower than for LIR protocol. On the other hand, LIR protocol has lower message delivery delay. For the scenario with relatively small number of sources (routes), we present modified protocols *mMDR* and *mLIR*. These protocols exploit the advantage of having only limited number of routes $N_r$ which are simultaneously active in the network, resulting in lower interference level.

The *mMDR* protocol, if possible, reduces the scheduling cycle from 7 slots to $T_{\min}^{(2)}$ which is necessary to provide scheduling for all transmissions where the interfering distance $d_i$ is larger than a given threshold $d_r$. For the *mLIR* protocol, it is not necessary to check simultaneous availability of $N/K$ terminals when searching for $k_0$ type of user but only $N_r$ terminals. These protocols are used for concrete scenarios to generate the results presented in the sequel.

Based on the previous explanations, the performance of *modified Minimum Distance Routing (mMDR)* and *modified Limited Interference Routing (mLIR)* protocols is shown by using the topology in Fig. 12. In this topology, we assume that there are 6 sources of type $k = 1$, and a set of unavailable users that are marked with **x.** Their location is described in Table III (scenario 1). So, the users transmit by relaying to their adjacent users available until all transmissions reach the BS. The routes for the ideal case, when all users are available for relaying, are indicated with continuous arrows in Fig.12. The routes obtained by *mLIR* protocol for this scenario are indicated with dashed arrows. For *mLIR* protocol, users try to relay to the same type of adjacent user available $k_0$ common to all transmitters. For the scenario 1, $k_0 = 2$ as shown in Fig. 12. Later on, this scenario is modified to include different sets of unavailable users as shown in Table III.

Table III. Description of the scenarios

| scenario | unavailable users | rescheduling (mLIR) |
|---|---|---|
| 1 | $x=\{u^5(2,0^\circ),u^6(2,60^\circ),u^7(2,120^\circ),u^6(2,150^\circ),u^5(1,210^\circ),u^7(2,120^\circ),u^4(2,300^\circ)\}$ | $k_0 \to 2$ |
| 2 | $o=\{u^5(2,0^\circ),u^2(1,30^\circ),u^6(2,60^\circ),u^2(3,110^\circ),u^7(2,120^\circ),u^7(2,210^\circ),u^2(2,270^\circ)\}$ | $k_0 \to 3$ |
| 3 | $p=\{u^5(2,0^\circ),u^6(3,60^\circ),u^5(2,90^\circ),u^2(3,110^\circ),u^2(2,0^\circ),u^3(2,240^\circ),u^6(1,270^\circ),u^3(2,330^\circ)\}$ | $k_0 \to 7$ |
| 4 | All users type $k$=2 and 3 | $k_0 \to 5$ |
| 5 | $n=\{u^4(2,30^\circ),u^7(2,120^\circ),u^2(3,110^\circ),u^2(2,180^\circ),u^7(2,210^\circ),u^2(2,270^\circ),u^7(1,330^\circ),u^3(2,330^\circ)\}$ | $k_0 \to 6$ |
| 6 | $z=\{u^5(2,0^\circ),u^6(2,60^\circ),u^7(3,50^\circ),u^7(2,120^\circ),u^2(3,110^\circ),u^7(2,210^\circ),u^7(3,270^\circ),u^2(2,270^\circ),u^3(2,330^\circ),u^7(1,330^\circ)\}$ | $k_0 \to 4$ |

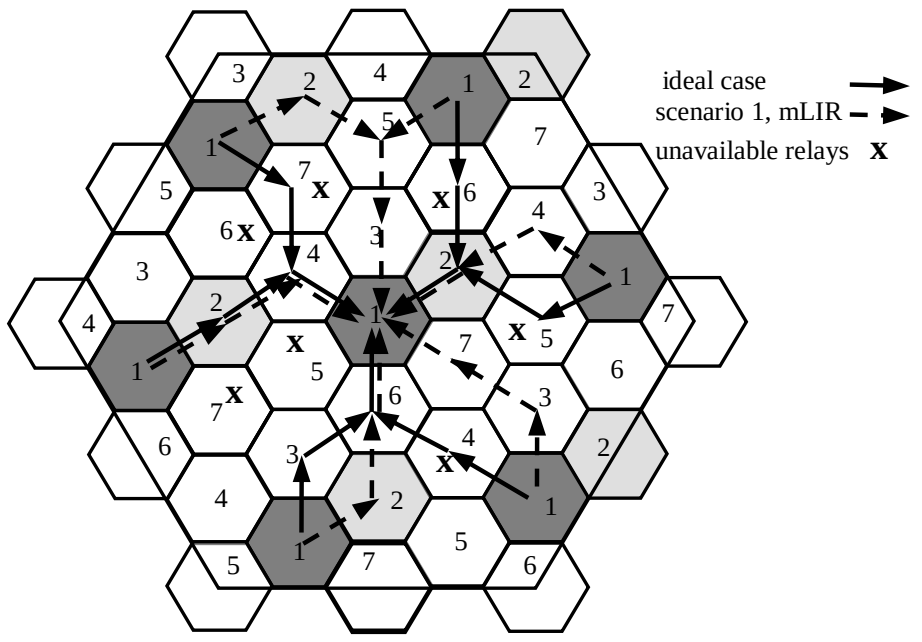

Fig. 12. Routing/scheduling scenario for $m^3$ network by using clustering factor $K$=7.

In Fig. 13 and 14, we present the capacity and throughput, respectively, versus the scenarios described in Table III for *mMDR* and *mLIR* protocols. The network capacity has been obtained as $C=\sum_{\Re i} C_{\Re i}$ where $C_{\Re i}$ is the route capacity obtained by (12). The throughput is given by $Thr = C/T$ where $T$ is the scheduling cycle. The results are compared to the ideal case when all users are available for relaying, and with another route discovery protocol referred to as LAR (Load Aware Routing). In LAR protocol, traffic load and power depletion are taken into account in the route discovery, so the protocol finds the route in such a way that the traffic is uniformly distributed through the whole network. In non ideal case, the highest capacity and throughput are obtained by *mLIR*. By *mMDR*, the users experience the shortest delay per route but, on the other hand, there is no control of the traffic distribution through the network. Consequently, there is more interference between adjacent links and the capacity is lower. The capacity obtained by LAR is larger than with *mMDR*. Although more slots are needed to complete the transmission with LAR, the gain obtained in distributing the traffic in some scenarios compensates the delay.

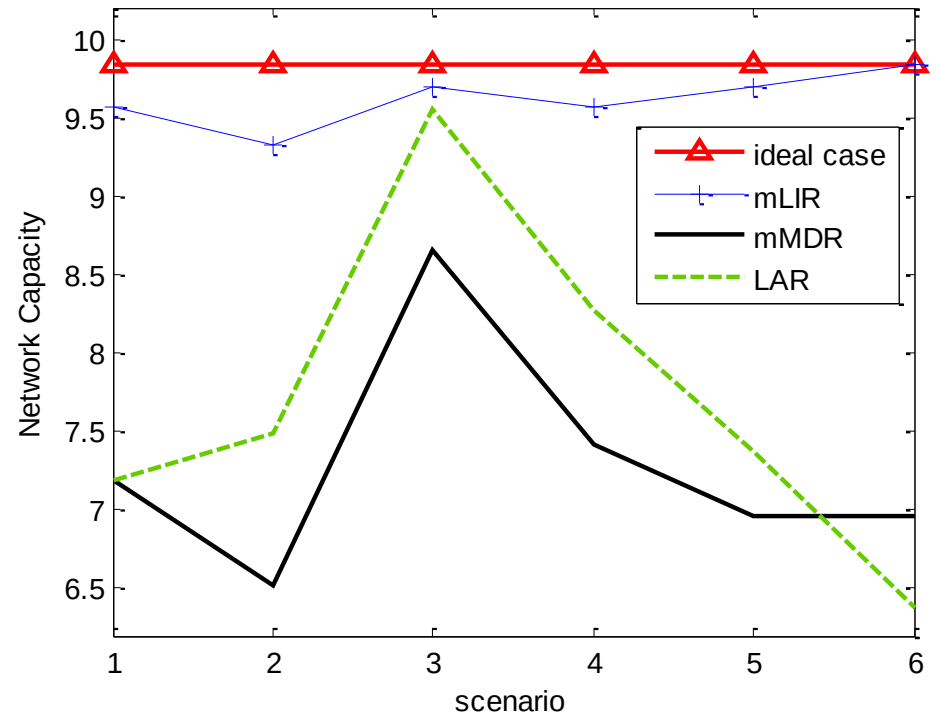

Fig. 13. Network Capacity.

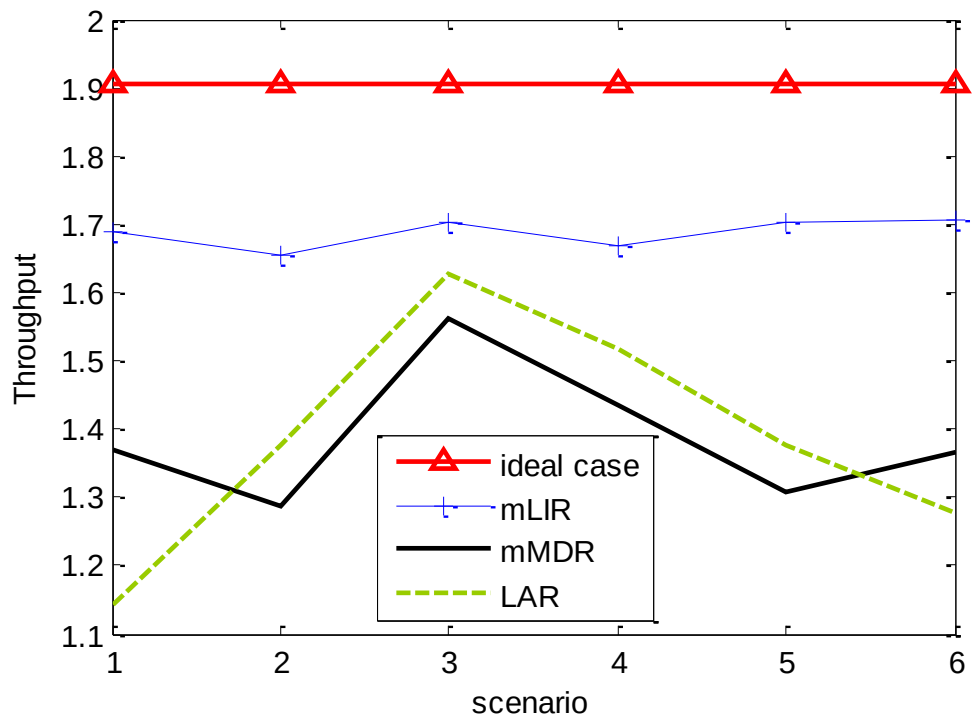

Fig.14. Throughput.

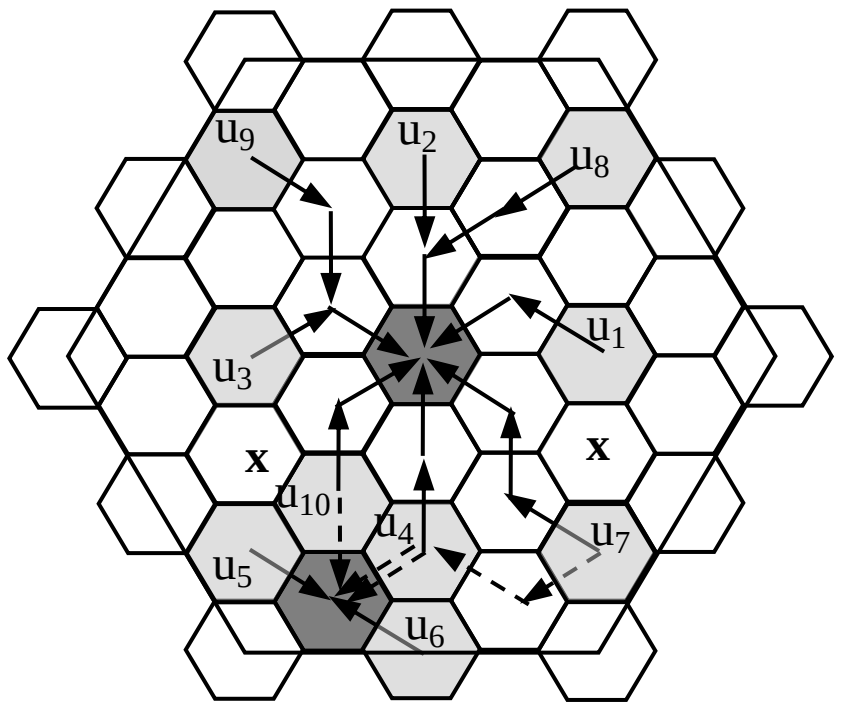

Fig. 15. $m^3$ network topology

### C) Traffic offloading incentives

We present some simulation results for a number of offloading scenarios where MNO and SSO cooperate to offload certain number of users through SSO. It is assumed that the availability probability is $p$=1. We consider the network topology shown in Fig. 15 and the scenarios described in Table IV. We assume that the coverage area of the WLAN is equal to the radius of the subcell, $2r$. The revenue of the MNO and SSO per unit of their respective utility functions is assumed to be $\rho = \rho_1 = 2$. In Fig. 16, we present results for the negotiation of the price $\chi$ between the MNO and SSO for the scenario 1 described in Table IV. ΔU and ΔU1 are the offset of the utility for the MNO and SSO, respectively, after and before offloading user 4. The optimum price $\chi$* obtained when there is equilibrium in the network (ΔU = ΔU1) is shown to be $\chi$* = 1.2. If a new user comes to the WLAN, $n_{\lambda wlan} = 1$, as described in scenario 2, the new price that the MNO will have to pay to the SSO for offloading user 4 is now decreased to $\chi$* = 0.8, as shown in Fig. 17. This is because the available capacity at the WLAN now is shared by one more user, so the utility for the offloaded user is now decreased and consequently, the price $\chi$* decreases, too. In scenario 3, a new user (user 7) transmits to the BS, $n_{\lambda bs} = 1$. The new price for offloading user 4, $\chi$* = 1.45, is obtained. As more users are now transmitting in the cellular network, the utility for MNO is decreased and there is more interest in offloading the user. The offset of the MNO, ΔU, is larger as shown in Fig. 16, so higher price can be paid for offloading (1.45 > 1.2). Instead, if we decide to offload user 7 through the WLAN (scenario 4), the equilibrium is obtained for $\chi$* = 3 as shown in Fig. 17. So, it would not payoff to offload this user as $\chi^* > \rho = \rho_1$. The utility for the SSO is reduced considerably as more slots are needed to complete the transmissions.

Table IV. Offloading scenarios as shown in Fig. 15

| scenario | MNO | SSO | offload |
|---|---|---|---|
| 1 | $u_1, u_2, u_3, u_4$ | $u_5$ | $u_4$ |
| 2 | $u_1, u_2, u_3, u_4$ | $u_5, u_6$ | $u_4$ |
| 3 | $u_1, u_2, u_3, u_4, u_7$ | $u_5$ | $u_4$ |
| 4 | $u_1, u_2, u_3, u_4, u_7$ | $u_5$ | $u_7$ |
| 5 | $u_1, u_2, u_3, u_4, u_7, u_8, u_9, u_{10}$ | $u_5$ | $u_4$ |
| 6 | $u_1, u_2, u_3, u_4, u_7, u_8, u_9, u_{10}$ | $u_5$ | $u_4, u_{10}$ |
| 7 | $u_1, u_2, u_3, u_4, u_7, u_8, u_9, u_{10}$ | $u_5$ | $u_4, u_7, u_{10}$ |

In scenario 5, the number of users transmitting to the MNO is increased now to 8 and the price obtained for offloading user 4 is $\chi$* = 1.18. As the number of transmissions in this scenario is rather high, the effects of offloading one user have less impact than for scenarios 1 and 3, so the price is lower. If we decide to offload one more user as shown in scenario 6, the offset obtained in the utility ΔU increases. Consequently, the price also increases to $\chi$* = 1.39 as the MNO has more interest in offloading. In scenario 7, we observe that the price for offloading also user 7 increases to $\chi$* = 2.05. As $\chi^* > \rho = \rho_1$, it would not payoff for the MNO to offload more users.

In Fig. 18, we consider scenarios 5, 6 and 7 again and we show how it affects to the optimum price $\chi$* to increase the number of users transmitting in the WLAN, $n_{\lambda wlan}$. As the available capacity at the WLAN is now shared by higher number of users, the capacity for the offloaded user decreases, which reduces the price $\chi$*. On the other hand, the price $\chi$* increases with the number of users offloaded $n_\mu$, as the offset ΔU is larger.

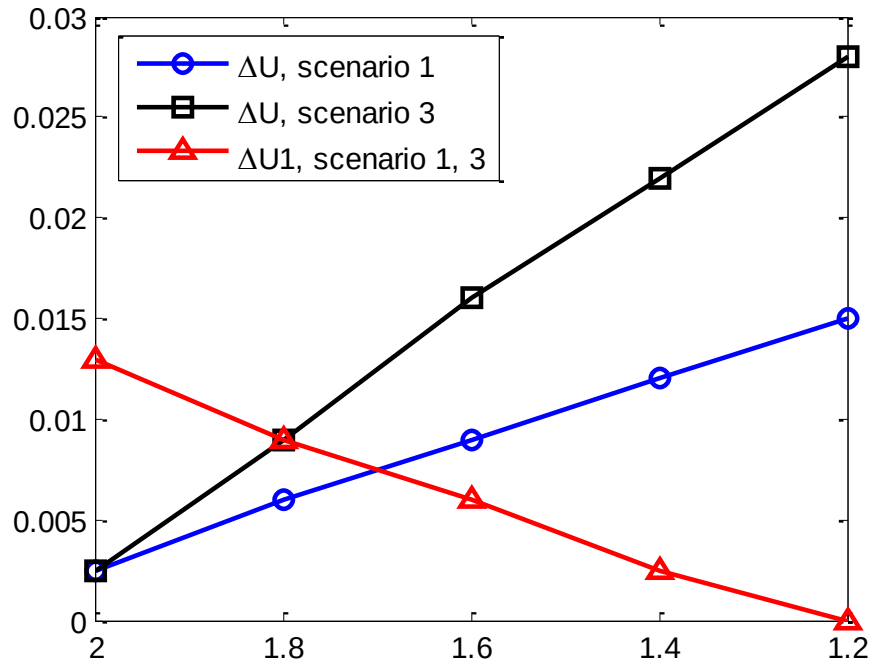

Fig. 16. ΔU and ΔU1 vs. $\chi$ for scenarios 1 and 3.

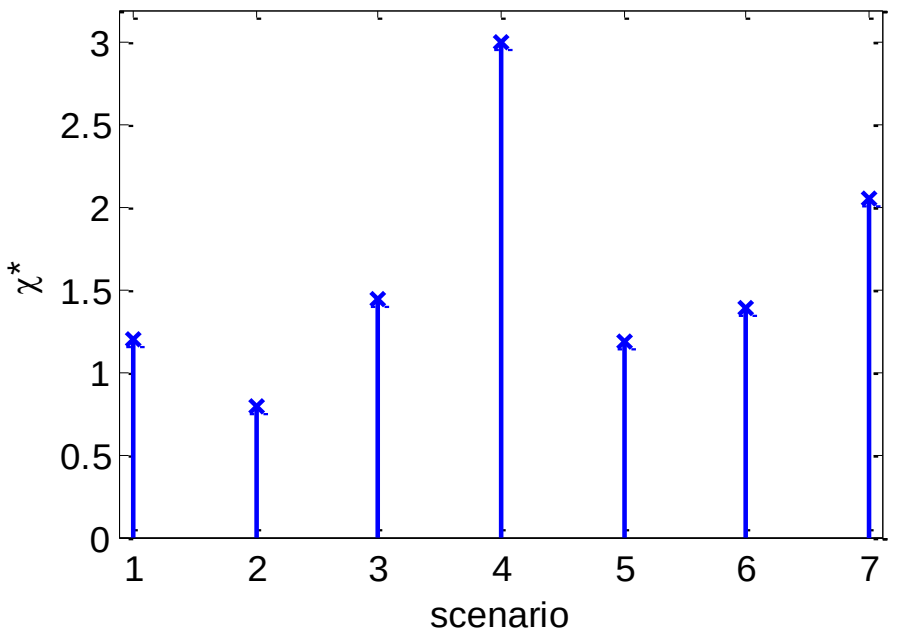

Fig. 17. Optimum price $\chi$* for scenarios 1-7.

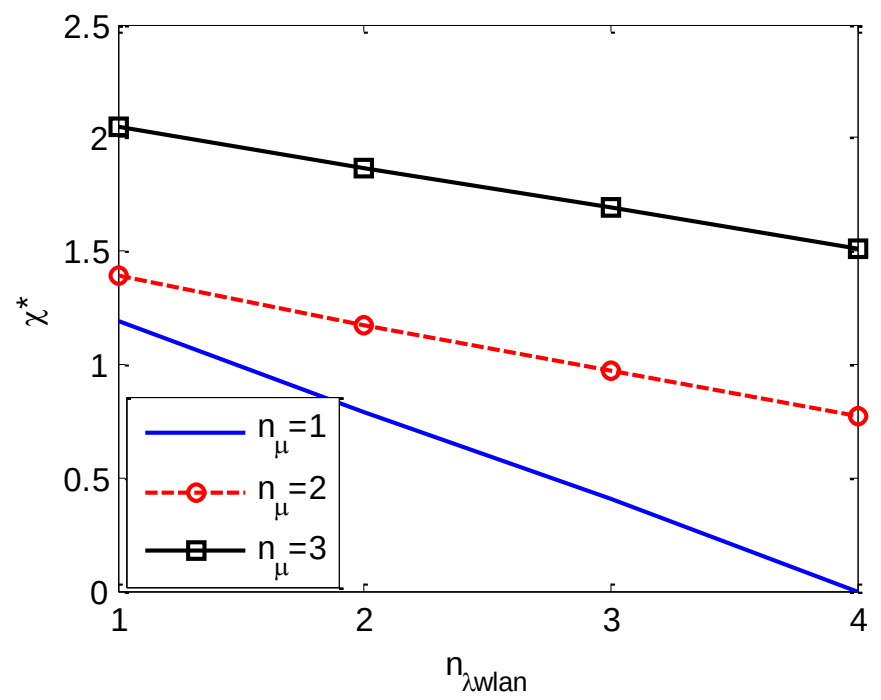

Fig. 18. Optimum price $\chi^*$ versus the number of new calls in the WLAN.

If we increase the WLAN coverage, under the assumption of proportional increase of the number of WLAN users, the offloading price will decrease since the WLAN capacity will be shared within more users. Besides, if the amount of allocated capacity to the offloaded users is still acceptable, the MNO will have interest to offload new users since it will result into larger remaining capacity for its own users.

*D) Implementation and impact of mobility*

As result of the terminals' mobility, the network will need to handle handoffs between the terminals and potential relays. The handoff mechanisms in our models are similar to those used in conventional networks with overlay of macro and micro/pico cells [28]. For this reason, we will not model these effects separately but just point out some solutions already used in practice. As a first step the traffic in the network should be classified, so that: a) Static, high data rate, delay tolerant traffic will be scheduled for multi-hop transmission with optimum tessellation ($H$=3 or 4) [29]. In this mode, high spatial resource reuse across the network can be achieved; b) The higher the mobility and the lower the delay tolerance, lower tessellation factor ($H$) should be used which guarantees lower number of handoffs in average; c) The users with highest mobility and the lowest delay tolerant traffic should transmit directly to the BS if the destination is not in the same macrocell. Otherwise, D2D option should be used. In this regime, resource reuse across the cell is low, if any; d) An interesting scenario arises when the terminal is forced (no other option is available) to relay the message to the terminal with different mobility. In this case, each terminal will be scheduled to operate with different tessellation factor ($H$).

The route discovery protocol is operated by the BS based on the terminal location information. All users communicate to the BS their position and willingness to cooperate on the conventional uplink signaling (control) channel of the macrocell. The position of the user is already tracked in the existing systems and only one additional bit (yes (1)/no (0)) is needed to transmit the information on willingness to cooperate. Once the position of the user is known to the base station the BS knows its relative position to the neighbors and in which order to run the route discovery protocol. The index of the current valid protocol (one bit for two options) and slot index (3 bits for 7 different options) for transmission are communicated back to the user on the existing downlink conventional signaling (control) channel. So, the additional overhead is negligible with respect to the capacity of the existing control channels used to set up the connection. The potential transmitter/receiver in the subcell is chosen to be the most static and centric (closest to the center of the subcell) user. The precise position of the potential relay is not important for the protocols which make them rather robust to positioning errors that are in the range of already existing technologies in cellular networks [30].

The same type of signaling is used between the MNO and SSO operator to exchange relevant information for the negotiation process (offloading price $\chi$ and set of users to be offloaded $\mathsf{N}_\mu$). We assume that the optimization processes (18) and (21) are solved fast enough to track the variation of the traffic in the network.

## VIII. Conclusion

In this paper, we present a comprehensive model to analyze the behavior of multi-hop, multi-operator, multi-technology ($m^3$) wireless networks which includes a number of relevant network parameters. The model compression techniques, mainly parameter aggregation and parameter absorption, are introduced to reduce the complexity of the optimization process. The model captures the interdependence between routing, scheduling and multi-operator incentive to cooperate when multiple technologies are available in a dynamic network. By making joint network access decisions, the utility of the Mobile Network Operator (MNO) and Small Scale Operator (SSO) are maximized.

Numerical results show that in a dynamic traffic environment the equilibrium price for traffic offload from cellular to WLAN network varies significantly. For the scenarios considered, this variation was by factor 3. It was also demonstrated that if the user availability is increased through the cooperation of multiple cellular operators, the network capacity can be increased up to 50% and the network throughput 30-40%.

The reduced transmission range enabled by multi-hop transmission makes the system transparent to mmWave technology which is a promising technology for next generation of cellular networks. For the application of mmWave, the physical layer model should be readjusted accordingly [24].